\documentstyle[prb,aps]{revtex}
\begin{document}
\draft
\title{Coherent Stranski-Krastanov growth in 1+1 dimensions with anharmonic 
interactions: An equilibrium study}
\author{Elka Korutcheva$^{1,*}$, Antonio M. Turiel$^{2}$ and Ivan 
Markov$^{3,**}$}
\address{$^{1}$Departamento de Fisica Fundamental, Universidad Nacional de 
Educacion a Distancia, 28040 Madrid, Spain}
\address{$^{2}$Laboratoire de Physique Statistique, Ecole Normal Superieure, 
24, rue Lhomond, 75231 Paris Cedex-05, France}
\address{$^{3}$Institute of Physical 
Chemistry, Bulgarian Academy of Sciences, 1113 Sofia, Bulgaria}
\date{\today}
\maketitle
\begin{abstract}
The formation of coherently strained three-dimensional islands on top of the 
wetting layer in Stranski-Krastanov mode of growth is considered in a model 
in 1+1 dimensions accounting for the anharmonicity and non-convexity of the 
real interatomic forces. It is shown that coherent 3D islands can be expected 
to form in compressed rather than in expanded overlayers beyond a critical 
lattice misfit. In the latter case the classical Stranski-Krastanov growth 
is expected to occur because the misfit dislocations can become energetically 
favored at smaller island sizes. The thermodynamic reason for coherent 3D 
islanding is the incomplete wetting owing to the weaker adhesion of the edge 
atoms. Monolayer height islands with a critical size appear as necessary 
precursors of the 3D islands. The latter explains the experimentally observed 
narrow size distribution of the 3D islands. The 2D-3D transformation takes 
place by consecutive rearrangements of mono- to bilayer, bi- to trilayer 
islands, etc., after exceeding the corresponding critical sizes. The 
rearrangements are initiated by nucleation events each next one requiring to 
overcome a lower energetic barrier. The model is in good qualitative 
agreement with available experimental observations.
\end{abstract}

\twocolumn
\section{Introduction}

The preparation of arrays of defect free three-dimen-sional (3D) 
nanoscale islands is a subject of intense research in the last decade owing 
to possible optoelectronic applications as quantum dots. The latter are 
promising for fabrication of lasers and light emitting 
diodes\cite{Seifert,Eagle,Leonard,Thanh}. In recent time the instability of 
two-dimensional (2D) growth against the formation of coherently strained 3D 
islands in highly mismatched heteroepitaxial systems has been successfully 
used to produce quantum dots. This is the well known Stranski-Krastanov (SK) 
growth mode where the decrement of the strain energy in the 3D islands 
overcompensates the contribution of the surface energy.

When the adhesion forces between the substrate and film materials 
overcompensate 
the strain energy stored in the overlayer owing to the lattice mismatch, a 
thin pseudomorphous wetting layer consisting of an integer number of 
monolayers is first formed by a layer-by-layer mode of growth. This kind of 
growth cannot continue indefinitely because of the accumulation of strain 
energy and the disappearance of the energetic influence of the substrate 
after several atomic diameters. Then, in the thermodynamic limit, unstrained 
3D islands are formed and grow on top of the wetting layer, the lattice 
misfit being accommodated by misfit dislocations (MDs) at the wetting layer 
- 3D islands boundary\cite{Kern,Markov1}. Thus the wetting layer and the 3D 
islands represent different phases in the sense of Gibbs\cite{Gibbs} 
separated by an interphase boundary. The energy of the latter is given by 
the energy of the array of MDs. This is the classical Stranski-Krastanov 
mechanism of growth\cite{Bauer} (see Fig. (\ref{SK}a)). However, it has 
been found that under certain conditions coherently strained (dislocation 
free) 3D islands are formed on top of the wetting layer (Fig. (\ref{SK}b)). 
These islands are strained to fit the wetting layer in the middle of their 
bases but are more or less strain-free near their top and side 
walls\cite{Ashu,Orr}. Such coherently 
strained islands are formed at large positive misfits when the lattice 
parameter of the overlayer is larger than that of the substrate and the 
overlayer is compressed. It has also been observed that the size distribution 
of the 3D islands is very narrow. The above observations have been reported 
for the growth of Ge on 
Si(100)\cite{Eagle,Thanh,Mo,Aumann,Voigt1,LeGoues,Ross,Kastner}, InAs on 
GaAs(100)\cite{Heitz,Koba,Moison,Yama,Joyce,Polim}, InGaAs on 
GaAs\cite{Leonard,Guha,Sneider,Temmyo}, and InP on 
In$_{0.5}$Ga$_{0.5}$P\cite{Carl}. In all cases the lattice misfit is positive 
and very large (4.2, 7.2, and $\approx $3.8\% for Ge/Si, InAs/GaAs, and 
InP/In$_{0.5}$Ga$_{0.5}$P, respectively) for 
semiconductor materials which are characterized by directional and brittle 
chemical bonds. The only exception to the authors' knowledge is the system 
PbSe/PbTe(111) in which the misfit is negative (-5.5\%) and the overlayer is 
expanded\cite{Pinc}. However, the authors of Ref. (\cite{Pinc}) note that 
whereas the in-plane 
lattice parameter of the PbSe wetting layer is strained to fit exactly the 
PbTe substrate, the parameter of the 3D islands rapidly decreases, reaching 
95\% of the bulk PbSe lattice constant at about 4 monolayers 
coverage\cite{Pinc}. One could speculate that the lattice misfit is 
accommodated by MDs introduced at the onset of the 3D islanding.

Whereas the classical SK growth is more or less clear from 
both thermodynamic and kinetic points of view, the formation of coherent 3D 
islands still lacks satisfactory explanation. We can consider as a first 
approximation the formation of coherent 3D islands in SK growth as {\it 
homoepitaxial growth on uniformly strained crystal surface both film and 
substrate materials having one and the same bonding}. If so, it is not clear 
what is the thermodynamic driving force for 3D islanding if the islands are 
coherently strained to the same degree as the wetting layer. It is also not 
clear why coherent 3D islands are observed in compressed rather than in 
expanded overlayers. 
Another question which should be answered is why the formation of coherent 
3D islands requires very large value of the positive misfit. The reason for 
the narrow size distribution is still unclear although much effort has been 
made to elucidate the problem\cite{Pr,Zang}. Finally, the mechanism of 
formation of coherent 3D islands is still an open question. 

Two major approximations are usually made when dealing theoretically with the 
formation of coherently strained 3D islands. The first is the use of the 
linear theory of elasticity in order to compute the strain contribution to 
the total energy of the 
islands\cite{Ashu,Orr,Pr,Voor,Chen,T1,T2,Rat,Sh,Daruka,Duport,S1,S2,Matt}. 
However, the validity of the latter is hard to accept bearing in 
mind the high values of the lattice mismatch. As will be shown below the 
MDs differ drastically in compressed and expanded films. Second, it is 
commonly accepted that the 
interfacial energy between the wetting layer and the dislocation free 
3D islands is sufficiently small and can be neglected in the case of coherent 
SK growth. The latter is equivalent to the assumption that the substrate 
(the wetting layer) wets completely the 3D 
islands\cite{Pr,Chen,T1,T2,Sh,Daruka,Duport,S1,S2}. In fact this assumption 
rules out the 3D islanding from thermodynamic point of view as 3D islands are 
only possible at incomplete wetting, or in other words, when the interfacial 
energy is greater than zero\cite{Bauer,Stranski1,Stranski2,Kaisch,Mark4}. As 
shown below the adhesion of the atoms to the wetting layer is 
also distributed along the island in addition to the strain distribution and 
plays a more significant role than the latter. Due to the lattice misfit the 
atoms are displaced from their equilibrium positions in the bottoms of the 
potential troughs they should ocupy at zero misfit. In such a way the 
adhesion of the atoms to the substrate is stronger in the middle of the 
islands and weaker at the free edges. The average adhesion of an island of a 
finite size is thus weaker compared with that of an infinite monolayer. An 
interfacial boundary appears and the wetting of the island by the substrate 
(the wetting layer) is incomplete in the average. It is this incomplete 
wetting which drives the formation of dislocation free 3D islands 
on the uniformly strained wetting layer. 

In the present paper we make use of a more realistic interatomic potential 
which is characterized by its anharmonicity, in the sense that the repulsive 
branch is steeper than the attractive branch, and by its nonconvexity, which 
means that it possesses an inflection point beyond which its curvature 
becomes negative. Recently Tam and Lam have used a Mie potential to describe 
the mode of growth in a kinetic Monte Carlo procedure\cite{Tan}. However, the 
above mentioned authors did not study the effect of misfit sign. Moreover, 
the distribution of the stress in the 3D islands has been studied again 
within the continuum elasticity theory\cite{Tan}. Yu and Madhukar\cite{Yu} 
computed the energy and the distribution of strain in coherent Ge islands on 
Si(001) using a molecular dynamics coupled with the Stillinger-Weber 
potential\cite{Still} but did not study the effect of anharmonicity in the 
general case.

The use of such a potential allows us to answer the question why coherently 
strained 3D islands appear predominantly in compressed 
overlayers. Comparing the energies of mono- and multilayer islands allows 
to make definite conclusions concerning the mechanism of formation and growth 
of the 3D islands, and the thermodynamic reason for the narrow size 
distribution. It turns out that there is a critical 2D island size above 
which the monolayer islands become unstable against the bilayer islands. 
Thus, as has been shown earlier by Stoyanov and Markov\cite{Stoyan,Markov1}, 
Priester and Lannoo\cite{Pr} and Chen and Washburn\cite{Chen}, the 
monolayer islands appear as necessary precursors for the formation of 3D 
islands. Beyond another critical size the bilayer islands become unstable 
against trilayer islands, etc. Then, the growth of 3D islands consists of 
consecutive transformations. As a result of each one of them the islands 
thicken by one monolayer. The critical size for the mono-bilayer 
transformations increases sharply with the decrease of the lattice misfit 
going asymptotically to infinity at some critical misfit. The monolayer 
islands are thus always stable against the multilayer islands below this 
critical misfit which explains the necessity of large misfit in order to 
grow coherent 3D islands. The critical misfit in expanded overlayers is 
nearly twice greater in absolute value than that in compressed overlayers 
which in turn explains why coherent 3D islanding is very rarely (if at all) 
observed in expanded overlayers.

The edge atoms are more weakly bound than the atoms in the middle of the 
islands. This is due to the weaker adhesion of the edge atoms to the wetting 
layer. Thus, the 2D-3D transformation takes place by transport of atoms from 
the edges of the monolayer islands, where they are weakly bound, on top of 
their surfaces to form islands of the upper layer where they are more 
strongly bound\cite{Markov1,Stoyan}. This process is then repeated in the 
transformation of bilayer to trilayer islands, etc. The critical size for 
the 2D-3D transformation to occur is the thermodynamic reason for the narrow 
size distribution of the 3D islands. 

In the case of expanded overlayers the atoms interact with each 
other through the weaker attractive branch of the potential and most of the 
atoms are not displaced from their equilibrium positions. The size effect is 
very weak, the average adhesion is sufficiently strong, and the critical 
sizes for 2D-3D transformation either do not exist or appear under extreme 
conditions of very large absolute value of the misfit. In any case MDs are 
introduced before the formation of bilayer islands. The coherent monolayer 
islands are either energetically stable against multilayer islands or MDs are 
introduced before the 2D-3D transformation. As a result the classical 
SK growth is expected in expanded overlayers. 

\section{Model}

We consider a model in 1+1 dimensions (substrate $+$ height) which we treat 
as a {\it cross-section} of the real 2+1 case. An implicit assumption is that 
in the real 2+1 case the monolayer islands have a compact rather than a 
fractal shape and the lattice misfit is one and the same in both orthogonal 
directions. Although the model is qualitative it gives correctly all the 
essential properties of the real 2+1 system as shown by Snyman and van der 
Merwe\cite{Sny1,Sny2,Sny3}. In this model the monolayer island is represented 
by a finite discrete Frenkel-Kontorova linear chain of atoms subject to an 
external periodic potential exerted by a rigid substrate 
(Fig. (\ref{frenkel}))\cite{Frenkel,FM1,FM2}. {\it We consider as a substrate 
the uniformly strained wetting layer of the same material consisting of an 
integer number of monolayers}. In other words, we consider the SK growth in 
two separate stages. The first stage is a Frank-van der Merwe 
(layer-by-layer) growth during which the wetting layer is formed. The second 
stage is a Volmer-Weber growth of 3D islands on top of the wetting layer. In 
this paper we restrict ourselves to the consideration of the second stage 
assuming the wetting layer is already built up. The energetic 
influence of the initial substrate is already lost and the bonding between 
the atoms in the 3D islands is the same as that of the atoms of the first 
atomic plane of the 3D islands to the atoms belonging to the uppermost 
plane of the wetting layer. 

The atoms of the chain are connected with bonds 
that obey the generalized Morse potential\cite{Mark2a,Mark2b,Mark3}
\begin{eqnarray}\label{potent}
V(x) = V_{o}\Biggl[\frac{\nu }{\mu - \nu }e^{-\mu (x-b)} - \frac{\mu }{\mu - 
\nu }e^{-\nu (x-b)}\Biggr],
\end{eqnarray}
shown in Fig. (\ref{potential}) where $\mu $ and $\nu $ ($\mu  > \nu $) are 
constants that govern the 
repulsive and the attractive branches, respectively, and $b$ is the 
equilibrium atom separation. For $\mu  = 2\nu $ the potential (\ref{potent}) 
turns into the familiar Morse potential. In the case of homoepitaxy the bond 
strength $V_{o}$ is related to the energy barrier for desorption.

The potential (\ref{potent}) possesses an inflection point 
$x_{inf} = b + \ln(\mu /\nu )/(\mu  - \nu )$ 
beyond which its curvature becomes negative. The latter 
leads to a distortion of the interatomic bonds in the sense that long, weak 
and short, strong bonds alternate (see the upper right-hand corner of Fig. 
(\ref{potential})\cite{Mark2a,Mark2b,Mark3,Haas}), and to the appearance of 
structures consisting of multiple MDs (multikinks or kink-antikink-kink 
solutions)\cite{Mark3}. The latter represent two kinks (or solitons) 
connected by a strongly stretched out bond (the antikink).

The 3D islands can be represented by linear chains stacked one upon the other 
as in the model proposed by Stoop and van der Merwe\cite{Stoop}, and by 
Ratsch and Zangwill\cite{Rat}, each upper chain being shorter than the 
lower one. In principle, the Frenkel-Kontorova model is inadequate to 
describe a thickening overlayer because of two basic assumptions inherent in 
it. The first one is the rigidity of the substrate. Assuming that the 
substrate remains rigid upon formation of 3D islands on top of it rules out 
the interaction between the islands through the elastic fields around them. 
It is believed that this assumption is valid for very thin deposits not 
exceeding one or two monolayers. The second one is connected with the 
relaxation effects. When a new monolayer island is formed on top of the 
previous one the latter should relax and the strains in the island will 
redistribute. One can expext that the formation, say, of a second monolayer 
will make the bonds between the first monolayer atoms effectively stiffer. 
As will be discussed below this will lead to weaker adherence of the atoms 
in the first monolayer to the wetting layer. MDs could be also introduced to 
relieve the strain. Nevertheless, the Frenkel-Kontorova model can provide 
excellent {\it qualitative} generalization in two dimensions both 
horizontally\cite{Sny1,Sny2,Sny3} 
and vertically\cite{Merwe2}. According to the authors of Ref. (\cite{Merwe2}) 
$n$-layer island can be mimicked by assuming that the force constant of the 
interatomic bonds is $n$ times greater than that of a monolayer island. Thus 
a bilayer island under compression could be simulated by doubling the value 
of the repulsive constant $\mu $. This approach obviously gives the upper 
bound of the effect of the next layers on the redistribution of the strain 
in the lower layers. An implicit shortcoming of this method is that it 
assumes the same number of bonds (and correspondingly atoms) in the upper 
chains and thus does not allow calculations of clusters with different 
slopes of the side walls.

Another approach to the problem has been proposed by Ratsch and 
Zangwill\cite{Rat}. They accepted that each layer (chain) presents a rigid 
sinusoidal potential to the chain of atoms on top of it. The atom, or more 
precisely, the potential trough separation of the lower chain is taken as 
average of all atom separations. As the strains of the bonds which are closer 
to the free ends are smaller, the average atom separation $b_{n}$ in the 
$n$th chain is closer to the unperturbed atom spacing $b$ and the lattice 
misfit $f_{n+1} = (b-b_{n})/b_{n}$ for computing the energy of the $n+1$st 
chain is smaller in absolute value than the misfit $f = (b-a)/a$ which is 
valid only for the base chain that is formed on the wetting layer the latter 
having an atom separation $a$. In such a way the 
lattice misfit and in turn the bond strains gradually decrease with the 
islands height. Every upper chain was taken shorter than the lower one by an 
arbitrary number of atoms and was centered on top of it as shown 
schematically in Fig. (\ref{3D}). Moreover, every uppermost chain is taken 
frozen (the relaxation of the lower chain upon formation of the next one is 
ruled out) and serves as a template for the formation of the next one. Then, 
the formation of each next chain does not exert any influence on the 
distribution of strain in the previous chains, and thus this approach 
represents the lower bound of the effect of the next layer.

In the present paper we will use the approach of Ratsch and 
Zangwill\cite{Rat}. The main reason is that it allows a gradual attenuation 
of the strain with the island height, and also different angles of the side 
walls. We believe that although rather crude this approach gives correctly 
the essential physics with one exception. It does not account for the 
decrease of the average adhesion of the base chain to the wetting layer upon 
thickening of the islands. An approximate evaluation of the latter effect can 
be obtained by using the upper bound approach. It should be emphasized that 
both approaches show {\it qualitatively identical} results. We could expect 
that the results of more accurate calculations including the strain 
relaxation will not differ qualitatively by those presented below. 
Preliminary studies with an energy minimization program allowing strain 
relaxation always produced dislocated expanded and coherent compressed 
islands in agreement with the results shown below. Note that owing to the 
approximations of the model (1+1 dimensions and the lack of relaxation) the 
figures obtained as a result of the calculations, e.g. 3.25\% for the 
critical misfit for 3D islanding, should not be taken as meaningful. Finally, 
we have to mention that the numerical solution of the system of governing 
equations (\ref{eqnset}) requires no more than a seconds on a 100 MHz PC 
even when the number of equations (atoms in the chain) is about 100. 

Discussing the stability of mono- and multilayer islands we follow the 
approach developed by Stoyanov and Markov\cite{Stoyan}. We start from the 
classical concept of minimum of the surface energy at a fixed volume. 
Following Stranski\cite{INS}, the surface energy $F(N)$ is defined as the 
difference between the potential energy of the cluster consisting of $N$ 
atoms and the potential energy of the same number of atoms in the bulk 
crystal
\begin{eqnarray}
F(N) = N\phi _{k} - \sum _{i=1}^{N}\phi _{i} \nonumber
\end{eqnarray}
which is valid for clusters with arbitrary shape and size. Here $\phi _{k}$ 
is the work necessary to detach one atom from a kink position (or the energy 
of atom in the bulk of the crystal), and the sum gives the work required to 
disintegrate the cluster into single atoms. Since the term $N\phi _{k}$ does 
not depend on the cluster shape the stability of mono- and multilayer islands 
is determined by the above sum. The maximum of the sum corresponds to a 
minimum of the surface (edge) energy of the cluster. Therefore, as a measure 
of stability, we adopt the potential energy per atom of the clusters, which 
is, in fact, equal to the above sum taken with a negative sign.

The potential energy of a chain of the $n$th layer consisting of $N_{n}$ 
atoms reads
\begin{equation}\label{energy}
E_{n} = \sum_{i=1}^{N_{n}-1}V(X_{i+1} - X_{i} - b) + 
\sum_{i=1}^{N_{n}}{\it \Phi} _{i}
\end{equation}
where
\begin{equation}\label{adhesi}
{\it \Phi} _{i} = 
\frac{W}{2}\Bigl[1-cos\Bigl(2\pi \frac{X_{i}}{b_{n-1}}\Bigr)\Bigr]
\end{equation}
accounts for the adhesion of the $i$th atom. $X_{i}$ are the coordinates of 
the atoms taken from an arbitrary origin. The difference $\Delta X_{i} = 
X_{i+1} - X_{i}$ is in fact the distance between the $i+1$st and $i$th atoms. 
The first sum in Eq. (\ref{energy}) gives the energy of the bond strains. The 
second sum gives the energy of the atoms in the periodic potential field  
created by the lower chain where $W$ is its amplitude and $b_{n-1}$ is the 
average potential trough separation  of the underlying layer. In general $W$ 
should be a function of the atom separation of the underlying layer and thus 
should depend on $n$ but for simplicity we neglect this dependence. As 
mentioned above $b_{n-1} = a$ holds only for the base chain. The amplitude 
$W$ can be considered in our model as the barrier for surface 
diffusion. On a nearest neighbor bond hypothesis $W$ is related to the 
substrate-deposit bond strength by
\begin{equation}
W = gV_{o}
\end{equation}
where $g < 1$ is a constant of proportionality varying approximately from 
1/30 for long-range van der Waals forces to 1/3 for short-range covalent 
bonds\cite{Merwe3}. 

The average of the second sum in Eq. (\ref{energy}) for the base chain 
divided by $V_{o}$ 
\begin{equation}\label{phi}
{\it \Phi}  = \frac{1}{N_{1}V_{o}}\sum_{i=1}^{N_{1}} {\it \Phi} _{i}
\end{equation}
has the same physical meaning as the adhesion parameter 
\begin{eqnarray}
{\it \Phi}  = \frac{\sigma  + \sigma _{i} - \sigma _{s}}{2\sigma } = 
1 - \frac{\beta}{2\sigma }\nonumber
\end{eqnarray}
which accounts for the incomplete wetting of the 3D islands by the substrate 
in heteroepitaxy ($\sigma $, $\sigma _{i}$ and $\sigma _{s}$ being the 
specific surface energies of the overlayer, the interface and the substrate, 
respectively, and $\beta $ being the specific adhesion energy)\cite{Bauer}. 
In the case of the classical SK growth the adhesion parameter is given by 
${\it \Phi } = \epsilon _{d}/2\sigma $ where $\epsilon _{d}$ is the energy of 
a net of MDs\cite{Mark7}. We have the case of complete wetting when 
${\it \Phi } \le 0$. The formation of 3D islands can obviously take place 
only when $0 < {\it \Phi } < 1$\cite{Bauer}.

Minimization of $E_{n}$ with respect to $X_{i}$ results in a set of governing 
equations for the atom coordinates in the form
\begin{equation}\label{eqnset}
e^{-\mu \epsilon _{i+1}} - e^{-\nu \epsilon _{i+1}} - e^{-\mu \epsilon 
_{i}} + e^{-\nu \epsilon _{i}} + A\sin(2\pi \xi _{i}) = 0,
\end{equation}
where $\epsilon _{i} = b_{n-1}(\xi _{i} - \xi _{i-1} - f_{n})$ is the strain 
of the $i$th bond, $\xi _{i} = X_{i}/b_{n-1}$ is the displacement of the 
$i$th atom with respect to the bottom of the $i$th potential trough, $f_{n}$ 
is the misfit between the $n$th chain and the substrate potential exerted by 
the $n-1$st chain, and $A = \pi W(\mu  - \nu )/\mu \nu b_{n-1}V_{o}$. The 
lattice misfit has its largest value $f = (b - a)/a$ only for the base chain 
in multilayer islands, and goes to zero with increasing islands thickness. 
Expanding of the exponentials in Taylor series for small strains gives the 
set of equations that govern the discrete harmonic model\cite{FM1,FM2}. 
Solving numerically the system of equations (\ref{eqnset}) gives the atom 
displacements $\xi _{i}$ and all the parameters characterizing the system 
can be easily computed. 

The properties of the solutions of the system (\ref{eqnset}) are of crucial 
importance for understanding of the coherent SK growth. Two forces act on 
each atom: first this is the force exerted by the neighboring atoms, and 
second, this is the force exerted by the substrate (the underlying chain or 
the wetting layer). The first force tends to 
preserve the natural spacing $b$ between the atoms, whereas the second force 
tends to place all the atoms at the bottoms of the corresponding potential 
troughs of the substrate separated at a distance $b_{n} \ne b$. As a result 
of the competition between the two 
forces the bond strains and the atom adhesion are distributed along the 
chain. The undislocated solution (Fig. (\ref{frenkel}a)) clearly shows the 
decrease of the atom adhesion at the ends of the chain as the atoms are more 
and more displaced towards the chain ends. In the case of positive misfit the 
dislocation represents an empty potential trough the bond in the core of the 
dislocation being strongly stretched out (Fig. (\ref{frenkel}b)). This 
picture is equivalent to a crystal plane in excess in the substrate. In the 
opposite case of negative misfit (Fig. (\ref{frenkel}c)) the dislocation 
represents two atoms in one trough (a crystal plane in excess in the 
overlayer), the bond in the dislocation core being compressed. Both 
configurations are energetically equivalent in the harmonic approximation 
where the force between the atoms increases linearly with the atom separation. 
This is not, however, the case when an anharmonic potential is adopted. The 
latter displays a maximum force between the atoms at $x = x_{inf}$. This is 
the theoretical tensile stress of the material 
$\sigma _{tens} = V_{o}\mu (\nu /\mu )^{\mu /(\mu  - \nu )}$ and if the 
actual force exerted on the corresponding bond is greater than $\sigma 
_{tens}$ the bond will break up\cite{Mark2a,Mark2b,Mark3,Mark11}. Thus the 
interval of existence of dislocated 
solutions in compressed chains depends on the material parameters $V_{o}, W, 
\mu , \nu , f$, and becomes very narrow. Dislocated solutions in compressed 
chains exist only in sufficiently long chains\cite{Mark2a,Mark2b,Mark3} 
beyond some critical chain length. As will be shown below, this leads to 
coherent SK growth in compressed overlayers. On the contrary, the bonds in 
the cores of the MDs in expanded chains are compressed and cannot break. As 
a result MDs become energetically favored and can be introduced in very 
short chains. Thus, the classical SK growth should be expected in expanded 
overlayers as the dislocated islands with a monolayer height can become 
energetically favored long before the coherently strained multilayer islands.

\section{Results}

\subsection{Monolayer islands}

The distribution of the bond strains along the chains is shown in Fig. 
(\ref{strain}a). As expected the bonds in the middle of the chains are 
strained to fit exactly the uniformly strained wetting layer. The strains 
at the chain ends tend to zero. In fact the strains of the hypothetic zeroth 
and $N$th bonds should be exactly equal to zero\cite{FM1,FM2}. The strains in 
the middle of the expanded chain compared with these of compressed ones are 
much closer to $-f$ owing to the weaker attraction between the atoms of the 
chain. Fig. (\ref{strain}b) 
shows the distribution of the bond energy. It is seen that in the case of 
compressed chains ($f > 0$) the bond energy in the middle of the chain is 
smaller than that in expanded chains owing to the stronger atom repulsion. 

The distribution of the adhesion of the separate atoms $\it \Phi _{i}$ (Eq. 
(\ref{adhesi})) (taken in terms of the bond energy $V_{o}$ as 
$({\it \Phi _{i}} - V_{o})/V_{o}$) is demonstrated 
in Fig. (\ref{adhesion}). The weaker adhesion at the chain ends, which is 
often overlooked in theoretical models, is due to the displacement of the 
atoms from the bottoms of the potential troughs (see Fig. (\ref{frenkel}a)). 
What is more important is that the atoms in the expanded chains adhere much 
more strongly to the wetting layer compared with the atoms in the compressed 
chains. 

Fig. (\ref{adpar}) shows the dependence of the mean adhesion parameter ${\it 
\Phi }$ (Eq. (\ref{phi})) on the number of atoms. As can be expected the 
atom adhesion in expanded overlayers is stronger than that in compressed 
ones owing to the weaker attraction between the atoms in the former. The 
forces exerted from the substrate are stronger than the forces between the 
chain atoms and the latter are 
situated more deeply in the potential troughs. The curves display maxima 
which are due to the interplay between the fraction of the most strongly 
displaced end atoms and the values of the particular displacements. 
In short chains the atoms are weakly displaced 
from the bottoms of the potential troughs and the adhesion is stronger. With 
increasing chain length the displacements of the end atoms increase and 
beyond some length saturate and do not increase anymore. The fraction of 
weakly displaced middle atoms increases and a maximum is displayed. The 
value of the maximum (not shown) decreases sharply with decreasing misfit 
going asymptotically to zero at zero misfit. This means that ${\it \Phi } > 
0$ at any value of the nonzero misfit which is the thermodynamic reason for 
3D islanding.

Fig. (\ref{totener}) shows the distribution of the total energy (strain plus 
adhesion) in chains with positive and negative misfit. The maxima in the 
middle are due to the strain contribution whereas 
the increase of the energy at the ends is due to the weaker adhesion. It is 
first seen that the atoms in the expanded chain are considerably more 
strongly bound to each other and to the substrate. The main difference 
between both curves is that the atoms at the free ends in compressed chains 
are much more weakly bound than the end atoms in expanded chains. This 
result is of crucial importance for our understanding of the mechanism of 
transformation of the mono- to multilayer (3D) islands. We conclude from Fig. 
(\ref{totener}) that compressed islands display a greater tendency to 
transform into bilayer islands and further to form coherent 3D islands in 
comparison with expanded islands.

\subsection{Multilayer islands}

The multilayer (3D) islands can be full or frusta of pyramids and can have 
side walls with different slope. The effect of the side walls slope on the 
minimum energy shape is more or less clear. More unsaturated 
dangling bonds normal to the film plane appear on side walls with smaller 
slope and the corresponding surface energy is greater. Obviously, the surface 
energy of the steepest walls with a slope of 60$^{o}$ is the lowest one. One 
could expect that the islands bounded with the steepest walls will be more 
stable than the flatter islands. The problem whether the pyramids are 
full or frusta is more difficult to resolve. First, with increasing 
pyramid height the lattice misfit decreases and the mean strain vanishes. 
This in turn leads to increase of the adhesion of the separate atoms, and, 
as a whole, to an increase of the bond energy closer to the apex of the 
pyramids. On the other hand, the layers which are closer to the apex are 
smaller in size and the size effect increases. The latter leads to smaller 
work of evaporation per atom of a whole uppermost atomic plane. As has been 
known for a long time, the work required to disintegrate a whole atomic plane 
into single atoms (the mean separation work) taken with a negative sign is 
equal to its chemical potential at the absolute zero\cite{Mark4,Kaischew}. 
Hence, adding to the pyramid smaller and smaller upper base atomic planes 
leads to a decrease of the mean separation work of the upper base and in turn 
to higher chemical potential. As a result we could expect that frusta of 
pyramids with a slope of $60^{o}$ of the side walls will be 
energetically favored. This is clearly seen in Fig. (\ref{pyrgrowth}) which 
demonstrates the energy per atom of pyramids with different side walls slopes 
as a function of the height taken as a number of monolayers. The curves 
display minima at a certain height which clearly show that the frusta of 
the pyramids are the lowest energy configurations. The energy of the full 
pyramids is much higher. The minimum of the $60^{o}$ side wall slope is the 
lowest one thus confirming the above consideration. The steepest side wall 
slope of $60^{o}$ is a natural consequence of the model which considers a 
face centered cubic rather than a diamond lattice. It is worth noting that 
Ratsch and Zanguill also report that the steepest side walls are 
energetically favored\cite{Rat}.

The above result does not mean that in the real experiment the coherent 3D 
islands will grow as frusta of pyramids. The lowest minimum in Fig. 
(\ref{pyrgrowth}) represents in fact the equilibrium shape of the islands. 
In reality the crystallites grow with a shape which is determined by the 
rates of growth of the different walls and thus depends on the 
supersaturation\cite{Chernov}. The growing crystal is bounded by the walls 
with the lowest growth rate at the given supersaturation. Mo {\it et al} 
have established with the help of scanning tunneling microscopy (STM) that 
small coherently strained Ge islands ("hut" islands) grow on Si(001) as 
full pyramids bounded with (105) side walls\cite{Mo}, whereas Voigtl\"ander 
and Zinner observed frusta of tetrahedron Ge pyramids on Si(111) with aspect 
(height-to-base) ratio showing a maximum of about 0.135 at a coverage of 4 
MLs\cite{Voigt1}. All the above is valid for sufficiently large crystals. We 
are interested here of the initial 
stages of growth of the 3D islands, or more precisely, in the transformation 
of monolayer into multilayer islands. As shown in the next sections the 
formation and growth of 3D islands proceeds by consecutive transformations of 
monolayer islands into bilayer and then into multilayer islands which is the 
lowest energy path of the 2D-3D transformation.

It should be stressed that the adhesion parameter $\it \Phi $ of a monolayer 
island should differ significantly from that of a multilayer island with the 
same base chain length. In our model they are equal. The reason is that the 
model does not allow the relaxation of the lower chains after formation of 
new ones on top of them. The above is obviously incorrect as the formation of 
a second chain on top of the base one leads to effectively stronger lateral 
bonds in the bilayer islands\cite{Merwe2}. We will try to qualitatively 
evaluate this problem and to discuss its consequences. As mentioned above the 
bilayer island could be treated as a first approximation as a monolayer 
island with a doubled force constant\cite{Merwe2}. As a result both the 
fraction of the strongly displaced end atoms and the corresponding 
displacements will be larger. Then 
the adhesion parameter of a bilayer island will be greater than that of a 
monolayer island with the same width. An evaluation of this effect 
can be made by using the approach of van der Merwe {\it et al} mentioned 
above\cite{Merwe2} by doubling of the constant $\mu $ in compressed 
chains. Thus for mono-, bi- and trilayer islands with $\mu  = 12, 24$ and 36,  
one obtains ${\it \Phi } = 0.024, 0.066$ and 0.1, respectively ($\nu  = 6, f 
= 0.05, N = 21$). As seen the effect of the third layer is weaker than that 
of the second which is easy to understand. The effect of formation of the 
next monolayers will have a smaller effect on the adhesion of the island and 
after some thickness the adhesion  parameter will not change anymore. Thus 
the base layer atoms in a coherent multilayer island are more weakly bound 
to the wetting layer. What follows is that once formed the bilayer islands 
stabilize the further growth of the coherent 3D islands.

\subsection{Stability of mono- and multilayer islands}

We compare further the energies of mono- and multilayer islands with different 
thickness. The latter 
are bounded with $60^{o}$ side walls as they have the lowest minimum energy 
as shown above. Fig. (\ref{rz1234+}a) shows the dependence of the 
energy of compressed monolayer and multilayer islands on the total number of 
atoms at comparatively small lattice misfit of 3\%. As seen the monolayer 
islands are always stable against bilayer and trilayer islands. A 2D-3D 
transformation is thus not expected and the film should continue to grow in 
a layer-by-layer mode coupled with an introduction of MDs at a later stage. 
The same dependence but at a larger 
misfit of 5\% is demonstrated in Fig. (\ref{rz1234+}b). The monolayer islands 
become unstable against the bilayer islands beyond a critical island size 
$N_{12}$, the bilayer islands in turn become unstable against the trilayer 
islands beyond a second critical number $N_{23}$, etc. The curve denoted by 
MD represents the energy of a monolayer chain containing one MD. The latter 
begins at a large number of atoms ($N = 52$) because the bonds in the cores 
of the MDs break up for shorter chains. This is due to the fact that the 
force exerted on these bonds from the neighboring atoms is greater than the 
theoretical tensile stress of the film material $\sigma _{tens}$ as mentioned 
above. Curve 1 which represents the energy of undislocated monolayer chain is 
computed for clarity up to a number of atoms smaller than the number (52) at 
which the solutions of the dislocated chain appear. The reason is 
that the values of the energy are very close and the curves are 
undistinguishable for the eye. The energies of monolayer chains with and 
without MDs cross each other at about $N = 300$ (not shown) which means that 
coherent 3D islands are formed long before the introduction of MDs. Moreover, 
the dislocated chain with a monolayer height has an energy much higher than 
the energies of the undislocated multilayer islands. The latter clearly shows 
that the film "prefers" to grow as coherent 3D islands in which the gradual 
decrease of the strain energy overcompensates the surface energy rather than 
to introduce MDs in the first monolayer.

Fig. (\ref{multiener-}) demonstrates the same dependence as in Fig. 
(\ref{rz1234+}) but in expanded chains. The absolute value of the negative 
misfit is very large (-10\%). At absolute values of the misfit smaller than 
5.5\% (not shown) the behavior of the energies is the same as in Fig. 
(\ref{rz1234+}a). The energies of the coherent mono- and multilayer chains 
cross again each other at some critical numbers of atoms but the dislocated 
monolayer chain (denoted by MD) becomes energetically favoured noticeably 
before the coherent bilayer chain to become stable. The classical SK growth 
should take place in expanded overlayers.

Fig. (\ref{n12f}) shows the misfit dependence of the first critical size 
$N_{12}$ for both positive and negative misfits. As seen it increases sharply 
with decreasing misfit going asymptotically to infinity at some critical 
misfits denoted by the vertical dashed lines. The existence of a critical 
positive misfit for coherent SK growth to occur explains why a high mismatch 
epitaxy is required in order to grow coherent 3D islands. The critical misfit 
below which the expanded monolayer islands are always stable against 
multilayer islands is nearly twice larger in absolute value compared with the 
same quantity in compressed overlayers. Thus coherent SK growth in expanded 
overlayers could be observed at unrealistically large absolute values of the 
negative misfit.

We conclude that the classical SK growth or a 2D growth will be observed in 
the thermodynamic limit at small positive misfits and a coherent SK growth at 
misfits greater than a critical misfit. This result clearly explains why 
large positive misfit is required for the coherent SK growth to 
occur. The large positive misfit leads to large atom displacements and in 
turn to weaker adhesion. The physics is essentially the same as in the case 
of heteroepitaxial growth of 3D islands directly on top of the surface of 
the foreign substrate (Volmer-Weber growth)\cite{Stoyan}.

\subsection{Mechanism of 2D-3D transformation}

It is natural to assume that once the monolayer islands become unstable 
against the bilayer islands ($N > N_{12}$) the former should rearrange 
themselves into bilayer islands. As shown below the mono-bilayer 
transformation can be considered as the first step for building sufficiently 
high 3D crystallites. The mechanism of the mono-bilayer transformation is 
easy to predict having in mind that the edge atoms are more weakly bound than 
the atoms in the middle. The edge atoms can detach and difuse on top of the 
monolayer islands giving rise of clusters of the second layer. We consider 
first in more detail the transformation of a monolayer island (chain) with  
a length $N_{o} > N_{12}$ into bilayer island. For this aim we plot the 
energy $E(n)$ of an incomplete bilayer island which consists of $N_{o} - n$ 
atoms in the lower layer and $n$ atoms in the upper layer referred to the 
energy $E_{o}$ of the initial chain consisting of $N_{o}$ atoms as a function 
of the number of atoms $n$ in the upper layer. This is the curve denoted by 
1-2 in Fig. (\ref{barr}). As seen it displays a maximum at $n = 1$ after 
which $\Delta E_{n} = E(n) - E_{o}$ decreases upto the complete mono-bilayer 
transformation at which $n = (N_{o} - 1)/2$. 

Curve 1-2 in Fig. (\ref{barr}) has the characteristic behavior of a 
nucleation process. The cluster at which the maximum of $\Delta E$ is 
observed can be considered as the critical nucleus of the second layer. As 
shown in Ref. (\cite{Stoyan}) the mono-bilayer transformation is a real 
nucleation process when 2+1 heteroepitaxial Volmer-Weber model is considered, 
in other words, when the 3D islands are formed directly on top of the foreign 
substrate without the formation of an intermediate wetting layer. The 
chemical potential of the upper island at the maximum is exactly equal to 
that of the initial monolayer island, and the supersaturation with 
which the nucleus of the second layer is in equilibrium is equal to the 
difference of the energies of desorption of the atoms from the same and the 
foreign substrate, This is namely the driving force for the 2D-3D 
transformation to occur. The 1+1 
model is in fact one-dimensional and the nuclei do not exist in the 
thermodynamic sense because the length of a row of atoms does not depend on 
the supersaturation\cite{Mark4,Stran}. However, considering our 1+1 model as 
a cross-section of the real 2+1 case we can treat the curve 1-2 in Fig. 
(\ref{barr}) as the size dependence of the free energy for nucleus formation 
and growth. We would like to emphasize that in the 2+1 case the nucleus does 
not necessarily consist of one atom. Its size should depend on the lattice 
misfit, and in the real situation - on the temperature. The curves denoted by 
2-3 and 3-4 in Fig. (\ref{barr}) represent the energy changes of bilayer to 
trilayer islands, and of trilayer to fourlayer islands, respectively. As seen 
they behave in the same way and the work for nucleus formation (the 
respective maxima) decrease with the thickening of the islands. The latter 
means that the mono-bilayer transformation is the rate determining process of 
the total mono-multilayer (2D-3D) transformation. 

\section{Discussion}

\begin{figure} 
\caption{\label{SK}
Schematic representation of (a) the classical SK growth, and (b) the 
coherent SK growth. In the latter case the side walls are shown steeper to 
demonstrate the compression exerted by the wetting layer. The MDs in (a) 
are denoted by inverse T's.}\
\caption{\label{potential}
The pairwise potential of Eq. (1) with $\mu  = 12$, $\nu  = 4$ and $V_{o} = 
1$. The dashed vertical line through the inflection point $x_{i}$ separates 
the regions of distortion ($x > x_{i}$) and undistortion ($x < x_{i}$) of 
the chemical bonds shown in the upper part of the figure.}\
\caption{\label{3D}
Schematic view of multilayer islands with different slopes of the side walls 
(a) 60$^{o}$, (b) 30$^{o}$, and (c) 19.1$^{o}$.}\
\caption{\label{frenkel}
Illustration of the solutions of the one-dimensional model of Frank and van 
der Merwe: (a) a chain without a misfit dislocation, (b) a misfit dislocation 
in a compressed chain, (c) a misfit dislocation in an expanded chain. With 
increasing the chain length in (a) the end atoms are more displaced from the 
bottoms of the potential troughs and approach the crests between them.}\
\caption{\label{strain}
Distribution (a) of the strain $\epsilon _{i} = \xi _{i+1} - \xi _{i} - f$ in 
a monolayer height compressed ($f = 0.07$) and expanded ($f = - 0.07$) chains, 
and (b) of the corresponding bond energy in units of $V_{o}$. 
$W/V_{o} = 1/3$, $\mu  = 12$, $\nu  = 6$.}\
\caption{\label{adhesion}
Distribution of the adhesion energy ${\it \Phi }_{i}/V_{o} - 1$ in 
monolayer height compressed (curve 1) and expanded (curve 2) chains. 
$W/V_{o} = 1/3$, $\mu  = 12$, $\nu  = 6$.}\
\caption{\label{adpar}
The mean adhesion parameter ${\it \Phi }$ as a function of the number of 
atoms in the chains for positive ($f = 0.07$) and negative ($f = - 0.07$) 
values of the misfit. $W/V_{o} = 1/3$, $\mu  = 12$, $\nu  = 6$.}\
\caption{\label{totener}
Distribution of the total energy (strain plus adhe- sion) in units of $V_{o}$ 
in monolayer height compressed ($f = 0.07$) and expanded ($f = - 0.07$) 
chains. $W/V_{o} = 1/3$, $\mu  = 12$, $\nu  = 6$.}\
\caption{\label{pyrgrowth}
Energy per atom of pyramidal 3D islands in units of $V_{o}$ with different 
slopes of the side walls denoted by the figures at each curve as a function 
of their thickness in number of monolayers. The number of atoms $N_{1} = 19$ 
in the base chain is one and the same for all curves. The frustum of the 
pyramid with a slope of $60^{o}$ of the side walls and height of 9 monolayers 
represents the equilibrium shape. $W/V_{o} = 1/3$, $\mu  = 12$, $\nu  = 6$.}\

\caption{\label{rz1234+}
The dependence of the energy per atom on the total number of atoms in 
compressed coherently strained islands with different thickness in monolayers 
denoted by the figures at each curve: (a) $f = 0.03$, (b) $f = 0.05$. The 
curve denoted by MD in (b) represents the energy of a monolayer chain 
containing one misfit dislocation. The numbers $N_{12}$, $N_{23}$, etc.  
give the limits of stability of monolayer, bilayer islands, respectively. 
$W/V_{o} = 1/3$, $\mu  = 12$, $\nu  = 6$.}\
\caption{\label{multiener-}
The dependence of the energy per atom on the total number of atoms in units of 
$V_{0}$ in expanded coherently strained islands with different thickness in 
monolayers de- noted by the figures at each curve, and at large negative value 
of the misfit $f = - 0.1$. The curve denoted by MD represents the energy of 
a monolayer chain containing one misfit dislocation. $W/V_{o} = 1/3$, 
$\mu  = 12$, $\nu  = 6$.}\
\caption{\label{n12f}
Misfit dependence of the critical size $N_{12}$. The vertical dashed lines 
denote the critical misfits below which $N_{12}$ is infinite. The curves are 
shown in one quadrant for easier comparison.}\
\caption{\label{barr}
The energy change $\Delta E_{n}$ in units of $V_{o}$ connected with the 
transformation of mono- to bilayer islands (curve 1-2), bi- to trilayer 
islands (curve 2-3),  and of tri- to fourlayer islands (curve 3-4), 
respectively, as a function of the number of atoms $n$ in the uppermost 
chain. $f = 0.05$, $W/V_{o} = 1/3$, $\mu  = 12$, $\nu  = 6$.}
\end{figure}
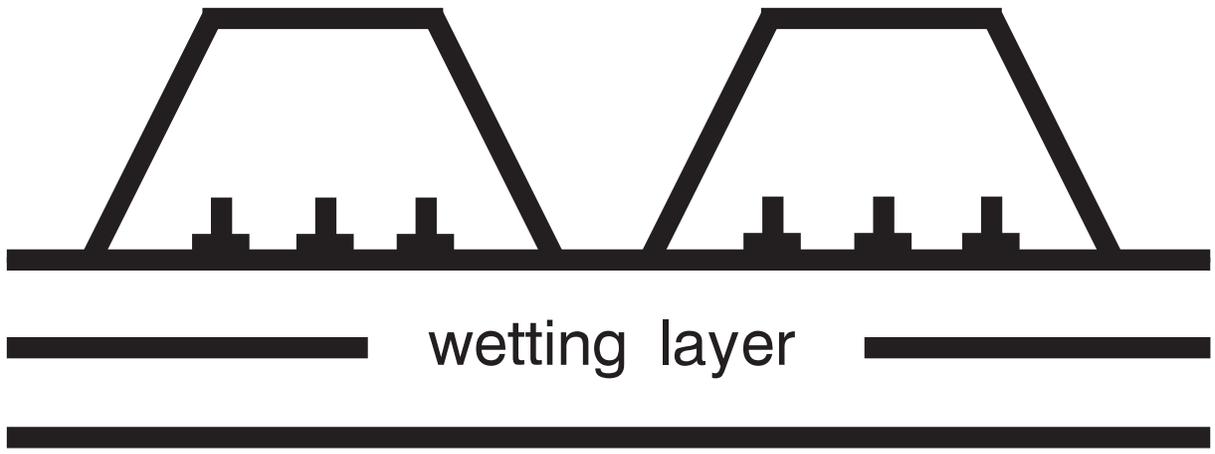
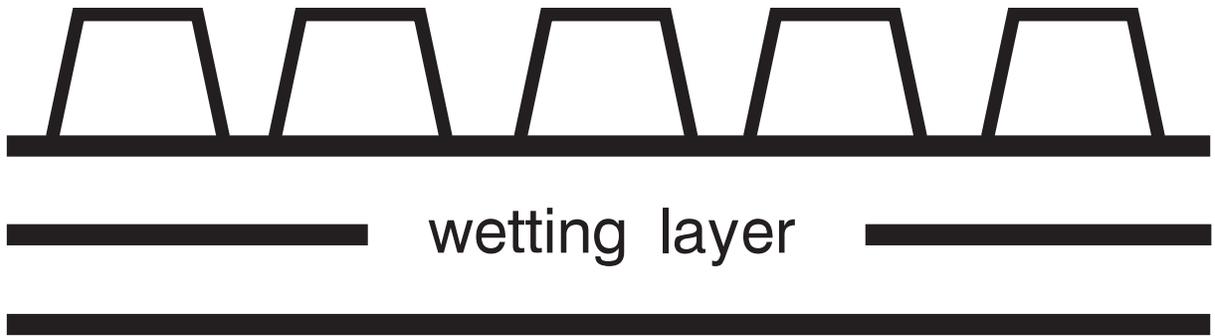
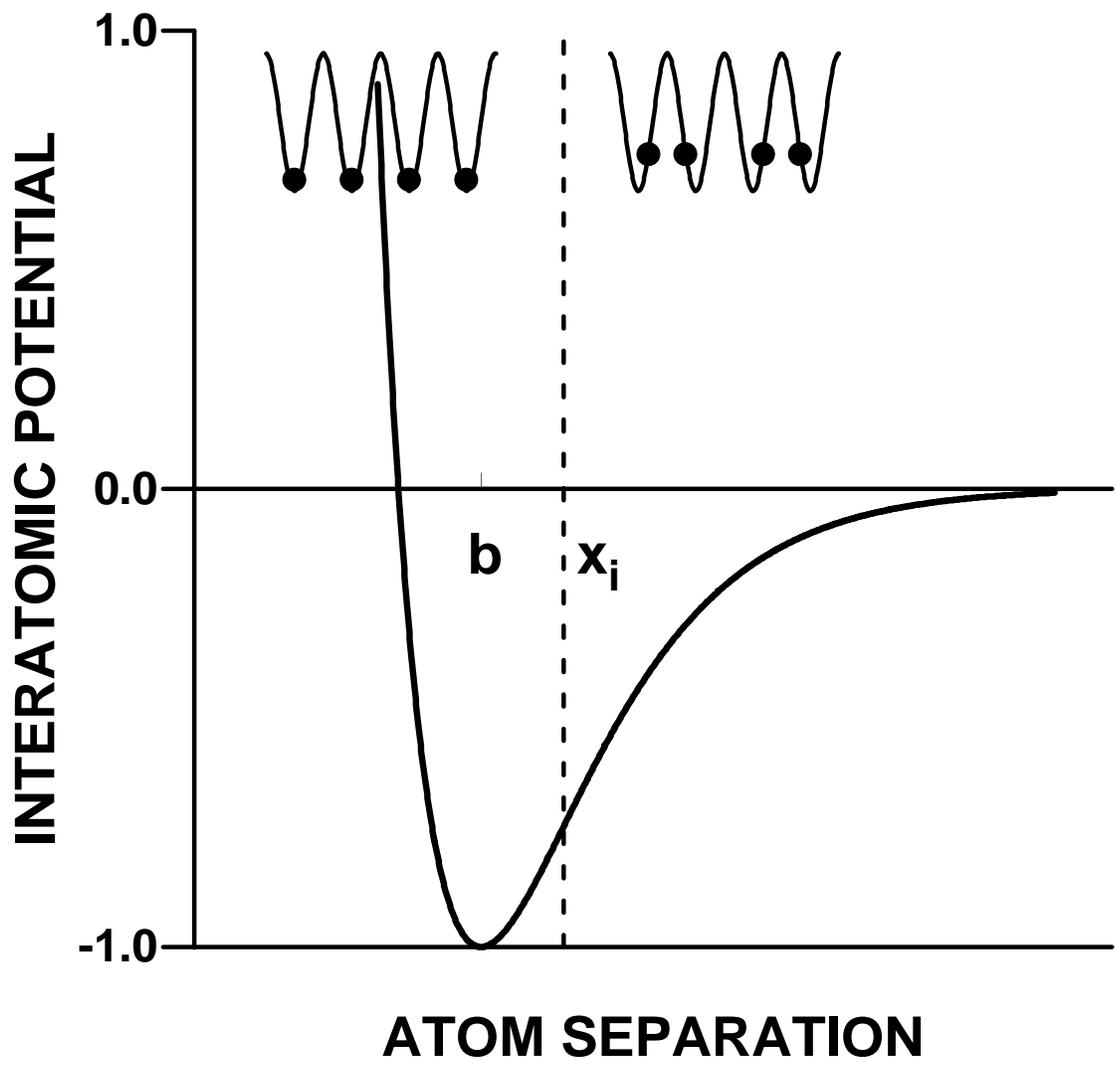
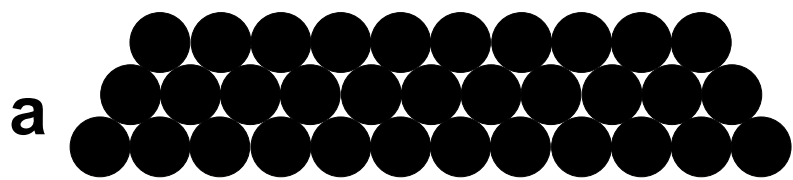
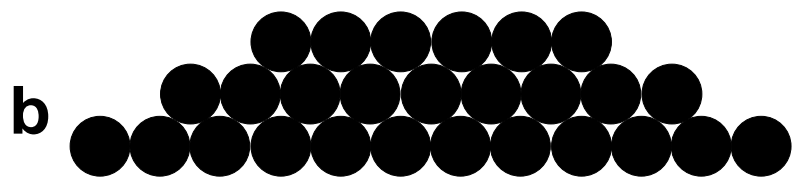
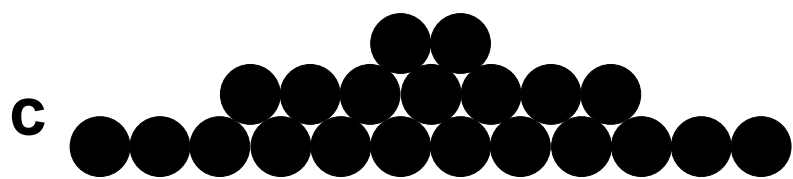
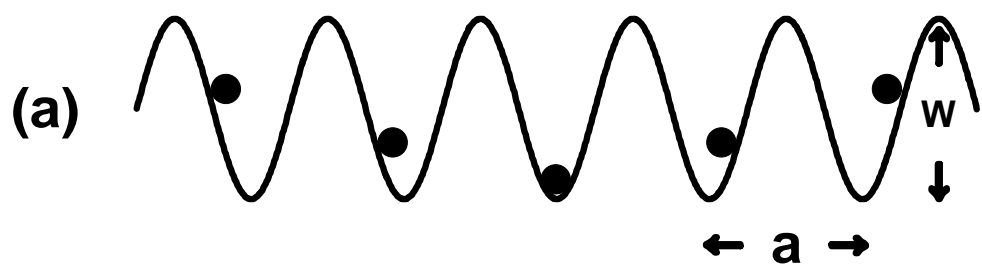
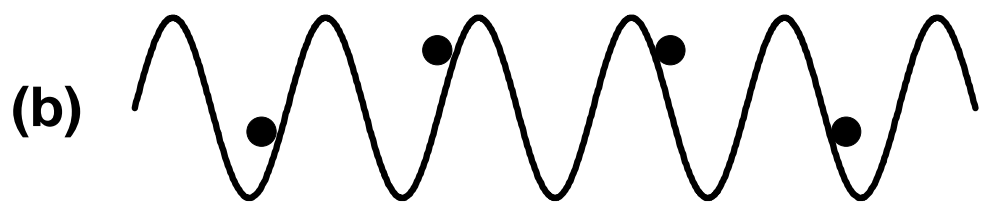
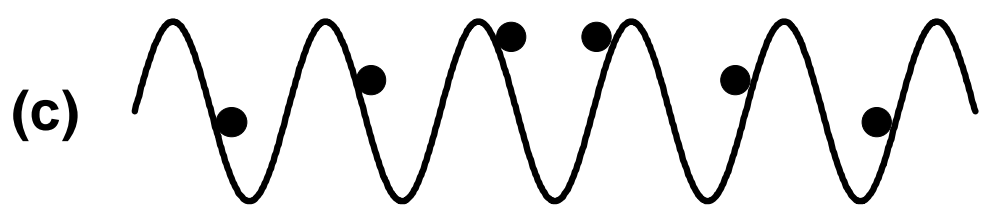
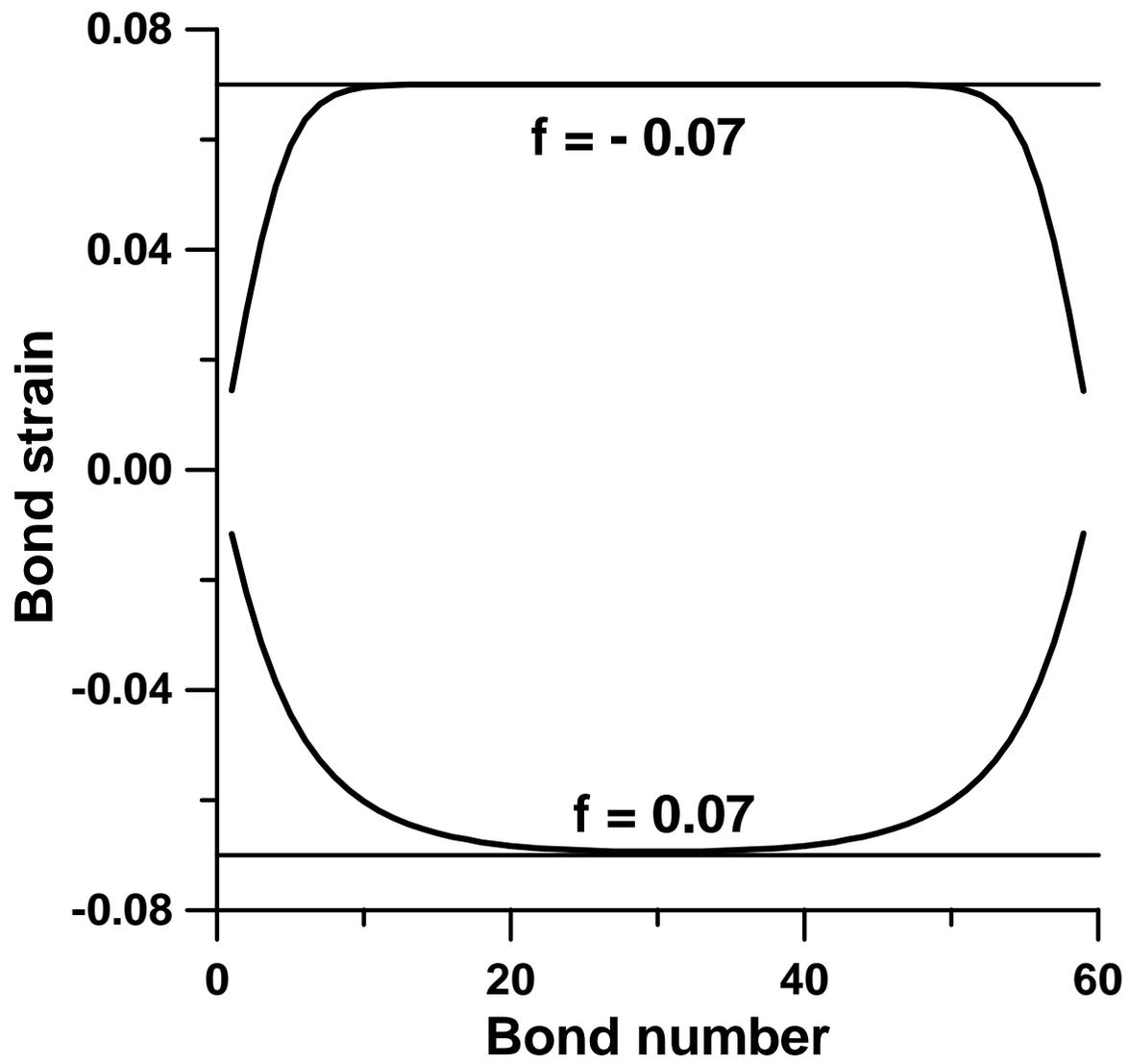
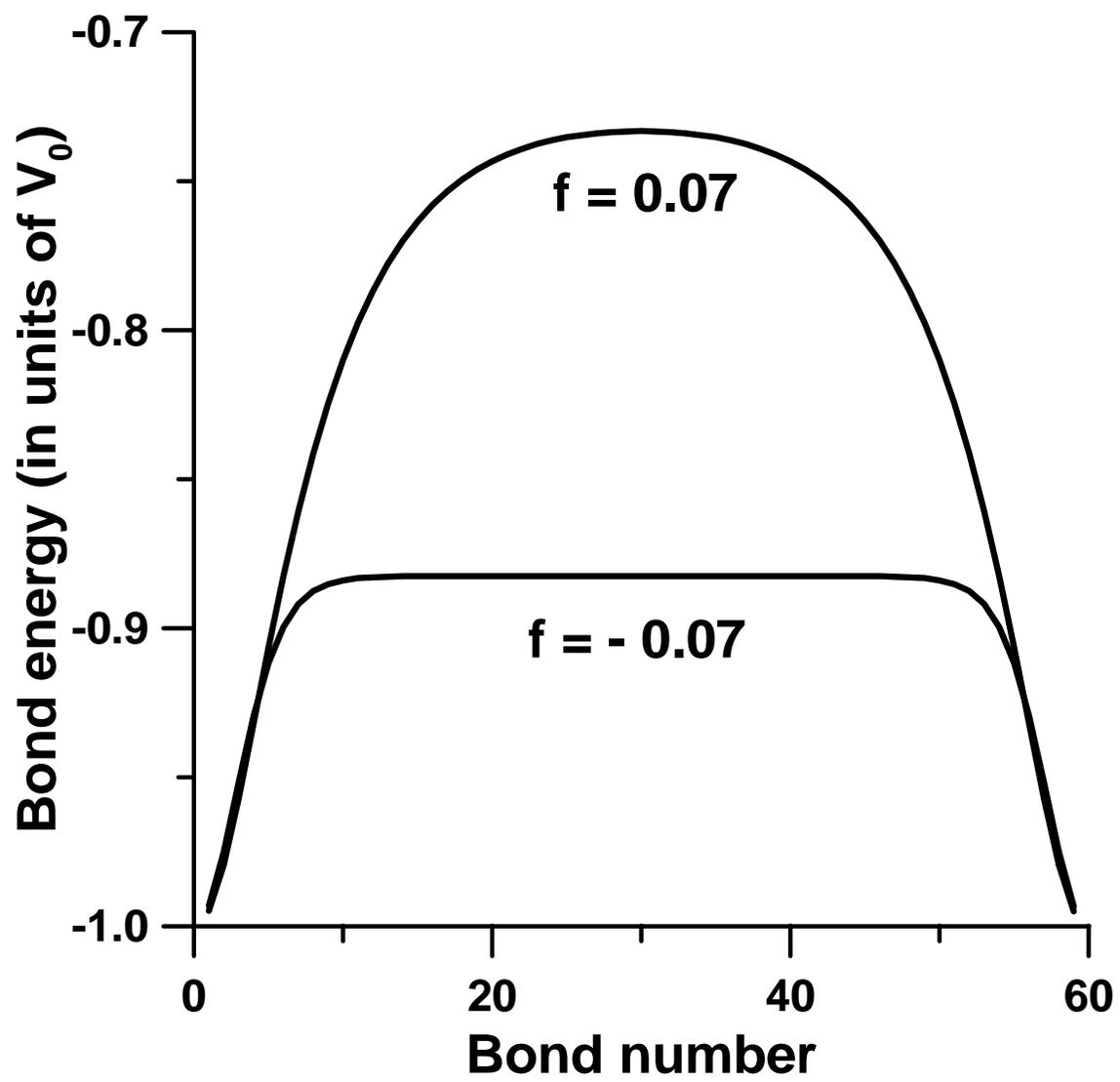
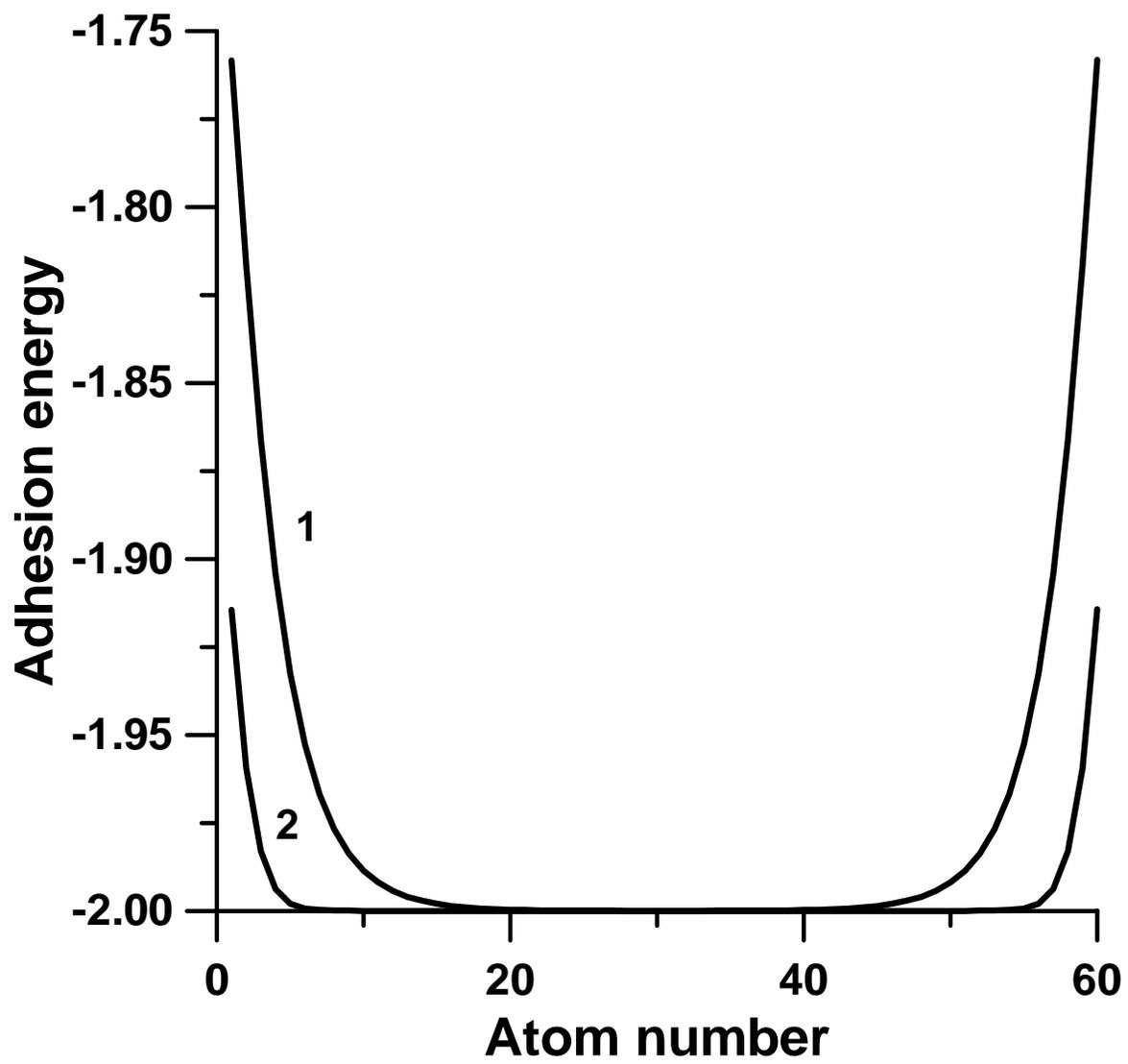
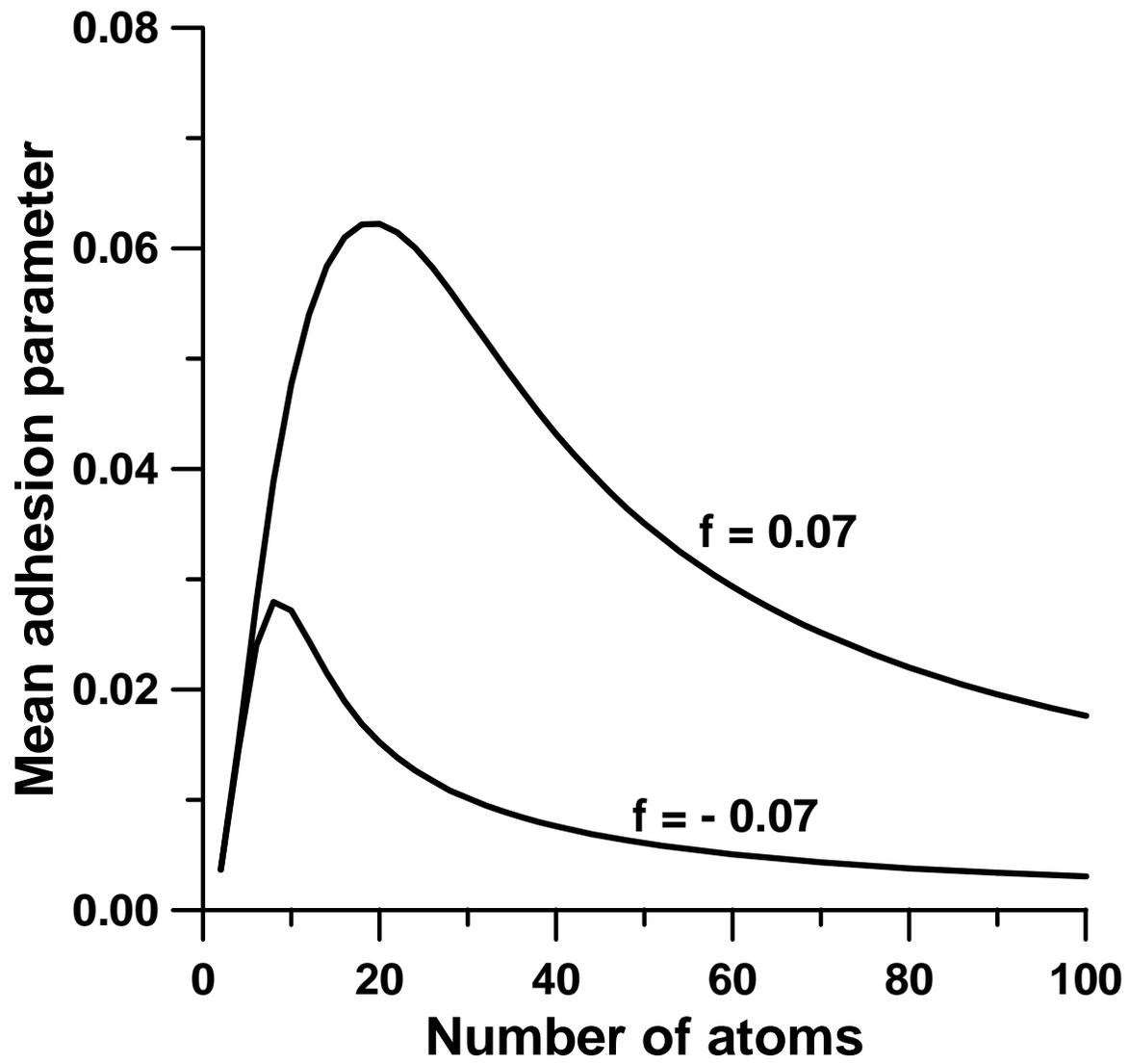
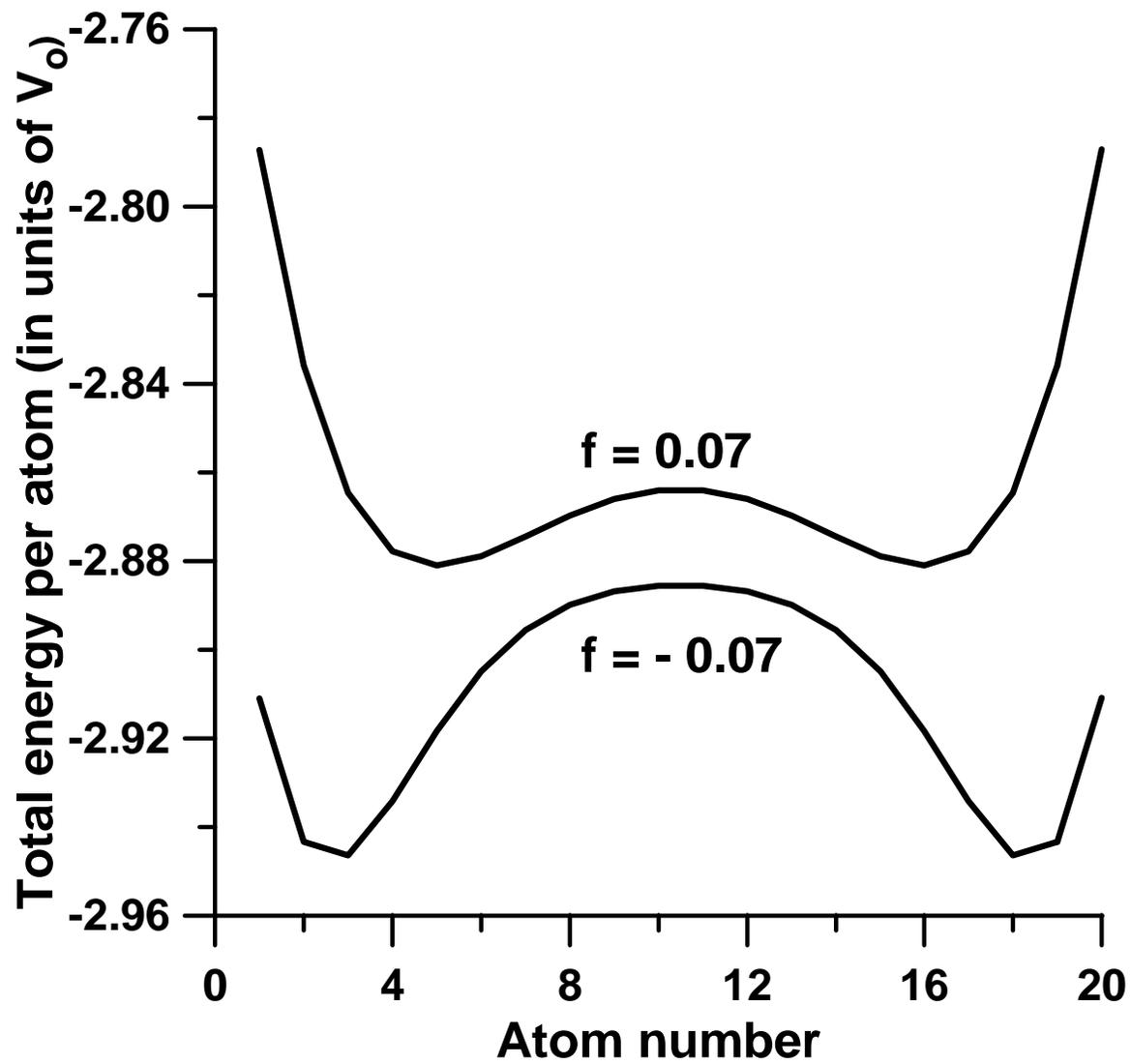
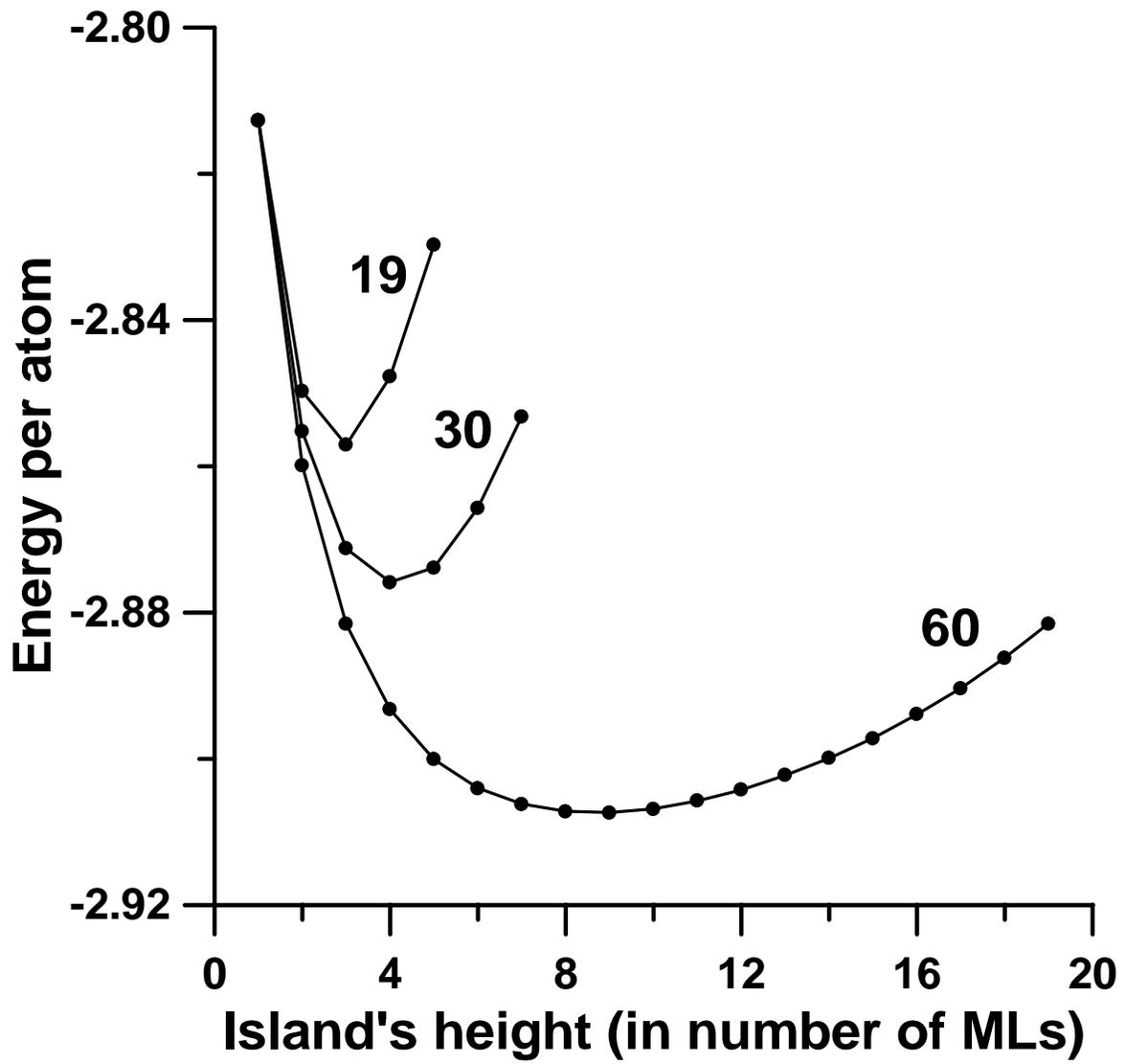
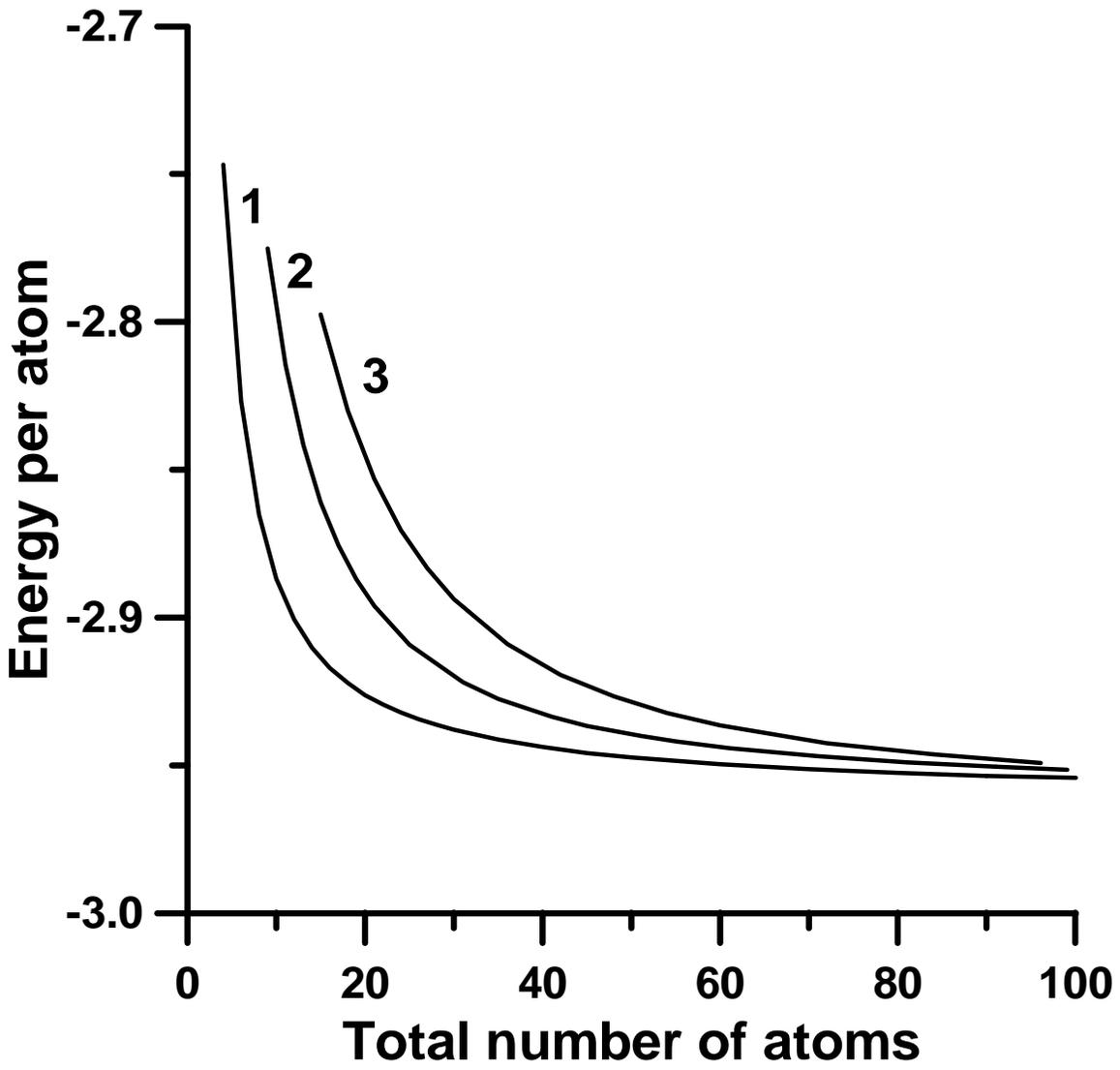
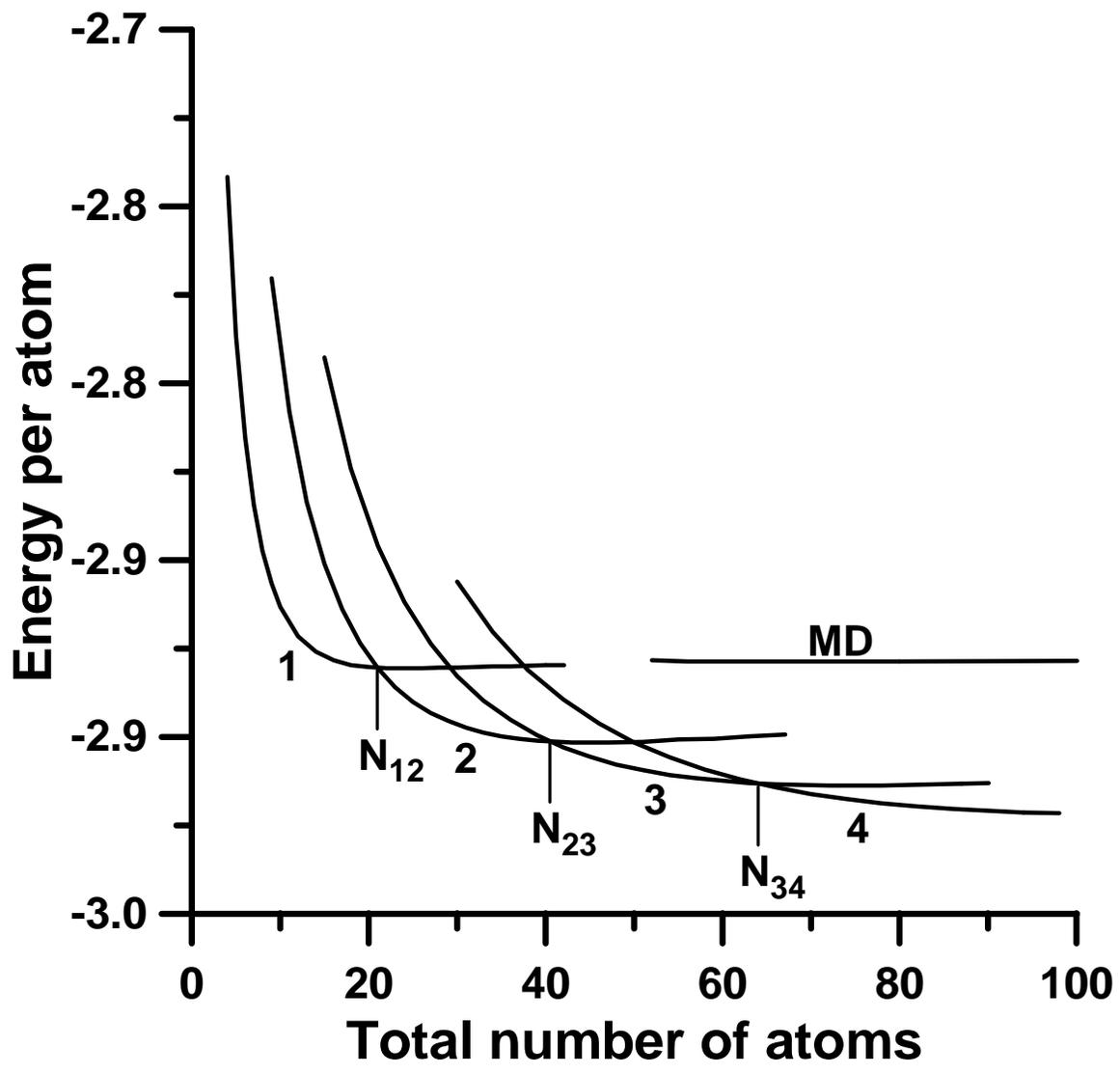
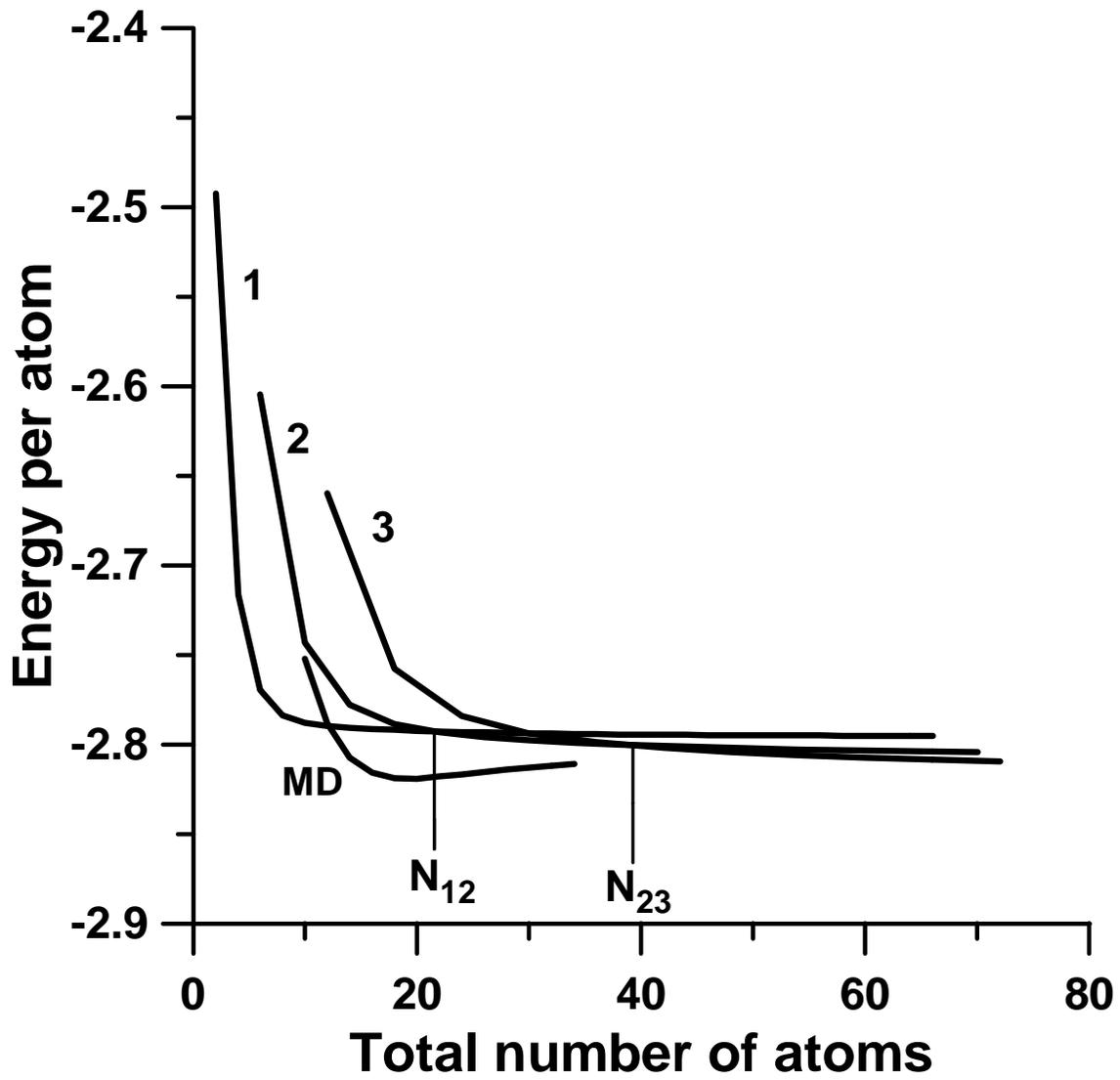
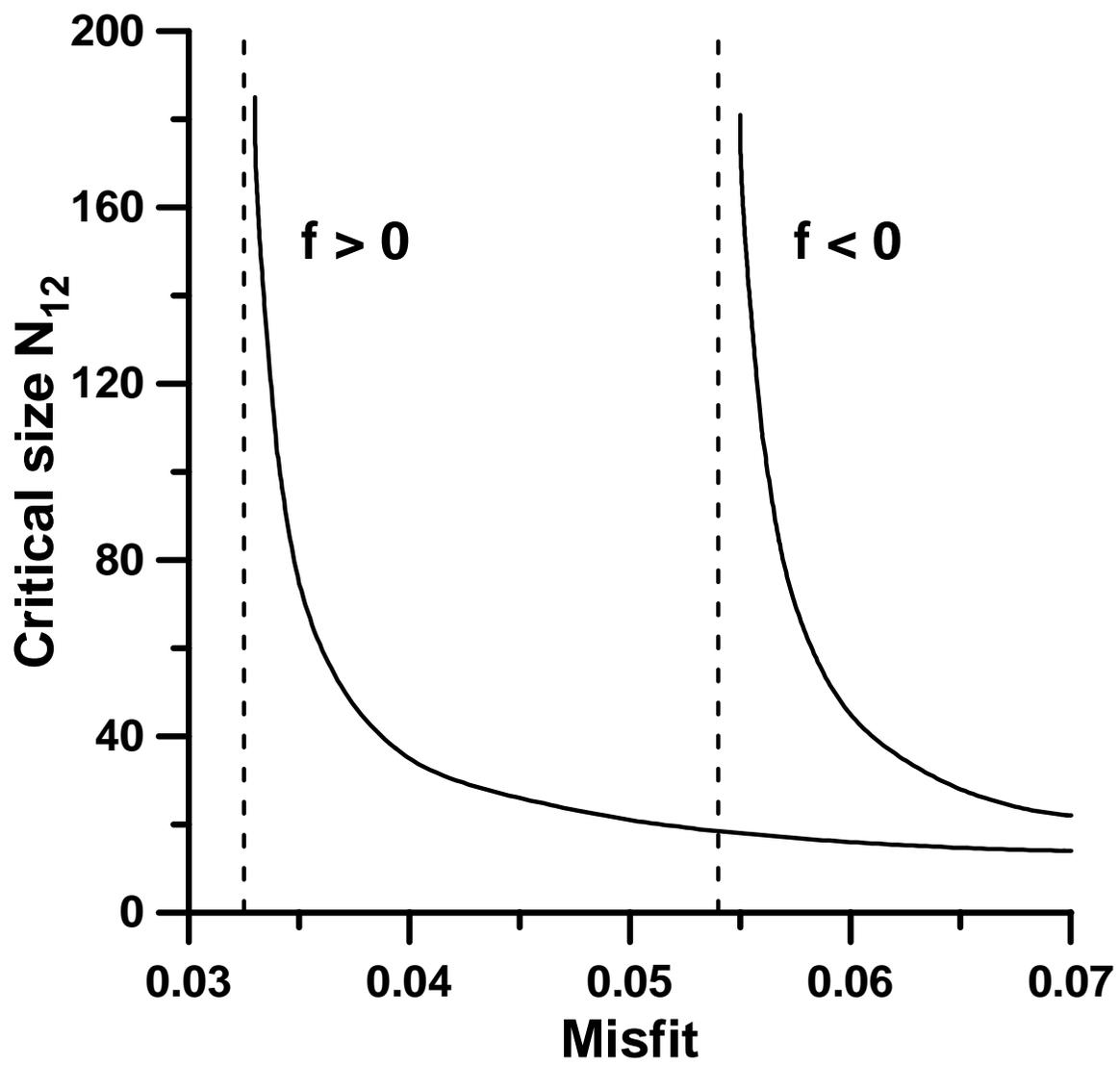
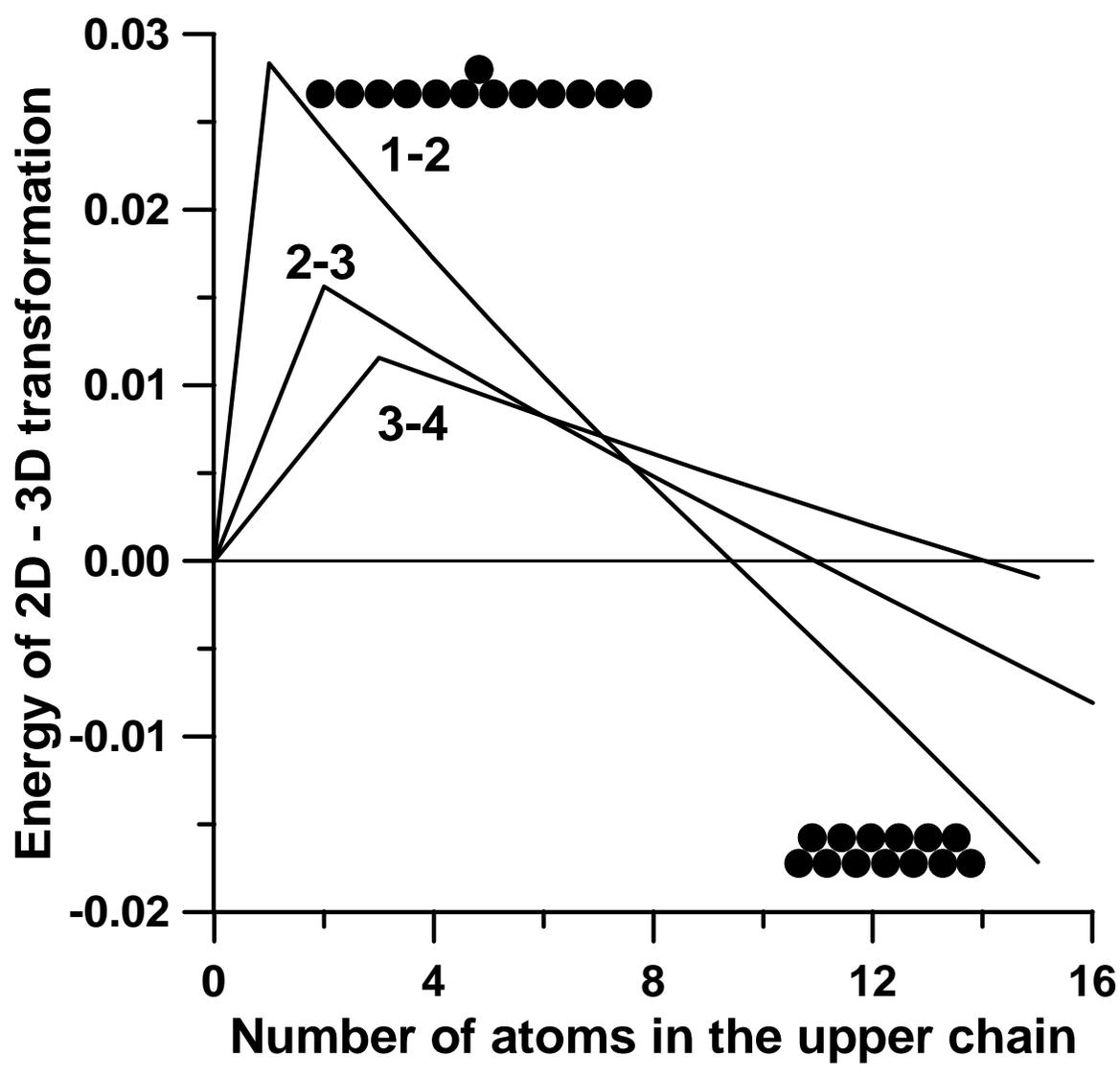

The Stranski-Krastanov growth mode appears as a result of the interplay of 
the film-substrate bonding, strain and surface energies. A wetting layer is 
first formed on top of which 3D islands nucleate and grow. The 3D islands and 
the wetting layer represent necessarily different phases. If this was not the 
case the growth should continue by 2D layers. Then we can consider as a 
useful approximation the 3D 
islanding on top of the uniformly strained wetting layer as Volmer-Weber 
growth. The latter requires the adhesion of the atoms to the substrate to be 
smaller than the cohesion between the overlayer atoms. In other words, 
the wetting of the substrate by the overlayer should be incomplete. In the 
classical SK growth this condition is fulfilled because of the formation of 
an array of misfit dislocations at the boundary between the islands and the 
wetting layer. The atoms are displaced from the bottoms of the potential 
troughs (mostly in the cores of the MDs, see Fig. (\ref{frenkel}b)) and 
(\ref{frenkel}c)) and thus are more weakly bound in average to the 
underlying wetting layer, irrespective of that the chemical bonding is one and 
the same. As a result the lattice misfit gives rise to an effective adhesion 
which is weaker than the cohesion of the overlayer atoms. Contrary to the 
wetting layer the 3D islands are elastically relaxed and their atom density 
differs from that of the former. Thus, the wetting layer and the 3D islands 
really represent different phases separated by a clear interfacial boundary 
whose energy is in fact the energy of the array of MDs. The physical reason 
for 3D islanding in the coherent SK growth is practically the same. In this 
case the atoms near the islands edges are displaced from the bottoms of the 
corresponding potential troughs (see Fig. (\ref{frenkel}a)) and they adhere 
more weakly to the wetting layer compared with the atoms in the middle. The 
thicker the islands the stronger is this tendency. Thus, the average adhesion 
of the 3D islands to the wetting layer is again weaker than the cohesion in 
the islands themselves. Then we can treat the coherent SK growth as a 
Volmer-Weber growth on top of the wetting layer. The main difference is that 
in Volmer-Weber growth the adhesion parameter $\it \Phi $ is constant 
whereas in the coherent SK mode it depends on the islands thickness. 

The weaker adhesion means in fact an  
incomplete wetting which appears as the thermodynamic driving force for the 
3D islanding. The smaller the misfit the smaller are the displacements of the 
edge atoms and in turn the stronger is the average wetting. The latter leads 
to the appearance of a critical misfit below which the edge effects do not 
play a significant role. The average wetting is very strong and the formation 
of coherent 3D islands becomes thermodynamically unfavored. The film will 
continue to grow in a 2D mode untill the strain is relaxed by introduction 
of MDs or dislocated 3D islands at a later stage. 
The existence of a critical misfit for 2D-3D transformation to occur both in 
compressed and expanded overlayers has been noticed in several studies. 
Pinczolits {\it et al}\cite{Pinc} have found that deposition of 
PbSe$_{1-x}$Te$_{x}$ on PbTe(111) remains purely two dimensional when the 
misfit is less than 1.6$\%$ in absolute value (Se content $< 30\%$). Leonard 
{\it et al}\cite{Leonard} have successfully grown quantum dots of 
In$_{x}$Ga$_{1-x}$As on GaAs(001) with $x = 0.5$ ($f \approx 3.6\%$) but 
60$\AA$ thick 2D quantum wells at $x = 0.17$ ($f \approx 1.2\%$). A critical 
misfit of 1.4\% has been found by Xie {\it et al} upon deposition of 
Si$_{0.5}$Ge$_{0.5}$ films on relaxed buffer layers of Si$_{x}$Ge$_{1-x}$ 
with varying composition\cite{Xie}.

The average adhesion (the wetting) depends strongly on the anharmonicity of 
the interatomic forces. Expanded islands adhere more strongly to the wetting 
layer and the critical misfit beyond which coherent 3D islanding is possible 
is much greater in absolute value compared with that in compressed overlayers. 
As a result coherent SK growth in expanded films could be expected at very 
(unrealistically) large absolute values of the negative misfit. The latter, 
however, depends on the materials parameters (degree of anharmonicity, 
strength of the chemical bonds, etc.) of the particular system and cannot be 
completely ruled out. Xie {\it et al}\cite{Xie} studied the deposition of 
Si$_{0.5}$Ge$_{0.5}$ films in the whole range of 2\% tensile misfit to 2\% 
compressive misfit on relaxed buffer layers of Si$_{x}$Ge$_{1-x}$ starting 
from $x = 0$ (pure Ge) to $x = 1$ (pure Si) and found that 3D islands are 
formed only under compressive misfit larger than 1.4\%. Films under tensile 
strain were thus stable against 3D islanding in excellent agreement with the 
predictions of our model.

The weaker average adhesion in compressed overlayers leads to another effect 
at misfits greater than the critical one. At some critical number of atoms 
$N_{12}$ the monolayer islands become unstable against the bilayer islands. 
The latter become in turn unstable against trilayer islands beyond another 
critical number $N_{23}$, and so on. As a result the complete 2D-3D 
transformation should take place during growth by consecutive transformations 
of mono- to bilayer, bi- to trilayer islands, etc. Owing to the stronger 
interatomic repulsive forces the edge atoms in the compressed monolayer 
islands adhere more weakly to the wetting layer compared with the edge atoms 
in expanded islands. This results in an easier transformation of mono- to 
bilayer islands which is the first step to the complete 2D-3D transformation. 
The latter includes also kinetics in the sense that the edge atoms have to 
detach and form the upper layers. However, it is not the strain at the edges 
(which is nearly zero) that is responsible for the easier detachment of the 
edge atoms as suggested by Kandel and Kaxiras\cite{Kandel} but the weaker 
adhesion. The 2D-3D transformation is hindered in expanded islands as the 
edge atoms adhere more strongly to the wetting layer. On the other hand, the 
existence of such critical sizes, which determine the intervals of stability 
of islands with different thickness, could be considered as the thermodynamic 
reason for the narrow size distribution of the 3D islands which is observed 
in the experiment. The latter does not mean that this is the only reason. 
Elastic interactions between islands and growth kinetics can have greater 
effect than the thermodynamics. The 2D-3D transformation takes place by 
consecutive nucleation events, each next one requiring to overcome a lower 
energetic barrier. Thus, the mono-bilayer transformation appears as the rate 
determining process. 

Let us consider all the above from another point of view. The results 
displayed in Fig. (\ref{rz1234+}b) show that the equilibrium shape aspect  
ratio increases gradually with the islands volume. The consecutive stability 
of islands with increasing thickness reflects the fact that the increase of 
the pyramid height is discreet (layer after layer) whereas the base chain  
length remains nearly constant. The stronger the adhesion or the smaller the 
misfit the wider will be the intervals of stability of islands with a fixed 
height and {\it vice versa}. The formation of every new crystal plane on the 
upper crystal face requires the appearance of a 2D nucleus. As the growing 
surface is usually very small the formation of one nucleus is sufficient for 
the growth of a new crystal plane. Thus we could expect a mononucleus 
layer-by-layer growth of  the pyramids\cite{Mark4,Chernov}. The latter has 
been independently established by using of a kinetic Monte Carlo method by 
Khor and Das Sarma\cite{das}. It should be noted that Duport, Priester and 
Villain established that the monolayer islands are thermodynamically favored 
upto a critical size beyond which the equilibrium shape becomes nearly a full 
pyramid\cite{Duport}. The transition from a monolayer island to a pyramid 
is of first order and requires the overcoming of an activation barrier which 
is proportional to $f^{-4}$.

It should be stressed that our definition of the critical 2D island size 
$N_{12}$ for 2D-3D transformation to begin differs from that in the papers 
of Priester and Lannoo\cite{Pr}, and of Chen and Washburn\cite{Chen}. The 
former authors define the critical size by comparing the energy per atom of 
monolayer islands with that of fully built 3D pyramids. Chen and Washburn 
have accepted as critical the size at which the energy of the monolayer 
islands displays a minimum\cite{Chen}. They found also that the critical size 
$N_{c}$ determined by the minimum of the energy increases very steeply with 
decreasing misfit ($N_{c} \propto f^{-6}$). Although our definition of 
$N_{c}$ is different we also observe a very steep misfit dependence (see Fig. 
(\ref{n12f})).

A rearrangement of monolayer height (2D) islands into multilayer (3D) islands 
has been reported by Moison {\it et al}\cite{Moison} who established that the 
InAs 3D islands begin to form on GaAs at a coverage of about 1.75 ML but then 
the latter suddenly decreases to 1.2 ML. This decrease of the coverage in the  
second monolayer could be interpreted as a rearrangement of an amount of 
nearly half a monolayer into 3D islands. The same phenomenon has been noticed 
by Shklyaev, Shibata and Ichikawa in the case of Ge/Si(111)\cite{Ichi}. 
Voigtl\"ander and Zinner noted that Ge 3D islands in Ge/Si(111) epitaxy have 
been observed at the same locations where 2D islands locally exceeded the 
critical wetting layer thickness of 2 bilayers\cite{Voigt1}.

Contrary to the linear theory of elasticity the anharmonicity and the 
non-convexity of the real interatomic potentials lead to different intervals 
of existence of misfit dislocations in compressed and expanded overlayers. 
The nonconvexity of the interatomic potential gives rise to the possibility 
of breaking of the expanded bonds in the cores of the MDs in compressed 
overlayers when the force exerted on them is greater than the theoretical 
tensile strength of the material. As a result MDs in compressed overlayers 
appear in sufficiently large islands and small coherent 3D islands can appear 
before that. On the contrary, this restriction does not exist in expanded 
overlayers where the bonds in the cores of the MDs are compressed. The 
introduction of MDs can thus become energetically favored in short chains 
(small islands) before the formation of coherent 3D islands and the classical 
SK growth should be observed in most cases. 

It should be noted that the results presented above depend on the 
approximations of the model particularly when the energy of the multilayer 
islands is computed. Allowing a strain relaxation of lower layers when new 
layers are formed on top of them could lead to earlier introduction of MDs 
but also to weaker adhesion of the 3D islands to the wetting layer. Thus, 
applying a more refined approach which accounts for the strain relaxation 
in the islands, as well as in the wetting layer, will allow us to study the 
transition from the coherent to the classical (dislocated) Stranski-Krastanov 
growth mode.

In summary, accounting for the anharmonicity and the non-convexity of the 
real interatomic potentials in a model in 1+1 dimensions, we have shown that 
coherent 3D islands can be formed on the wetting layer in the SK mode 
predominantly in compressed overlayers at sufficiently large values of the 
misfit. Coherent 3D islanding in expanded overlayers could be expected as an 
exception rather than as a rule. Monolayer height islands with a critical 
size appear as necessary precursors of the 3D islands. The latter explains 
the narrow size distribution of the 3D islands from thermodynamic point of  
view.

\acknowledgements

One of the authors (EK)  is financially supported by the Spanish DGES 
Contract PB97-0076 and partly by Contract F608 of the Bulgarian National Fund 
for Scientific Research. IM gratefully acknowledges the inspiration and the 
fruitfull discussions with R. Kaischew. The authors greatly benefitted from 
the remarks and criticism of Jacques Villain (Grenoble).

\end{document}